\documentclass{article}

\ifx\pdfoutput\undefined
\usepackage{graphicx}
\else
\usepackage[pdftex]{graphicx}
\usepackage{epstopdf}
\fi

\usepackage{amsmath,amssymb}
\usepackage{eucal}
\usepackage{euler}
\usepackage{xspace}
\usepackage{stmaryrd}

\usepackage{xy}
\xyoption{frame}
\xyoption{arrow}
\xyoption{curve}
\xyoption{arc}
\xyoption{line}
\xyoption{matrix}
\xyoption{tile}
\xyoption{ps}

\usepackage{amsthm}

\theoremstyle{plain}
    \newtheorem{theorem}{Theorem}
    \newtheorem{lemma}[theorem]{Lemma}
    \newtheorem{corollary}[theorem]{Corollary}
    \newtheorem{proposition}[theorem]{Proposition}
    \newtheorem{fact}[theorem]{Fact}
\theoremstyle{definition}

\theoremstyle{remark}
    \newtheorem{remark}[theorem]{Remark}

    \newtheorem{uclaim}{Claim}

\theoremstyle{plain}

\newcount\mscount
\newdimen\proofrulebreadth \proofrulebreadth=.05em
\newdimen\proofdotseparation \proofdotseparation=1.25ex
\newdimen\proofrulebaseline \proofrulebaseline=2ex
\newcount\proofdotnumber \proofdotnumber=3
\let\then\relax
\def\hfi{\hskip0pt plus.0001fil}
\mathchardef\squigto="3A3B
\newif\ifinsideprooftree\insideprooftreefalse
\newif\ifonleftofproofrule\onleftofproofrulefalse
\newif\ifproofdots\proofdotsfalse
\newif\ifdoubleproof\doubleprooffalse
\let\wereinproofbit\relax
\newdimen\shortenproofleft
\newdimen\shortenproofright
\newdimen\proofbelowshift
\newbox\proofabove
\newbox\proofbelow
\newbox\proofrulename
\def\shiftproofbelow{\let\next\relax\afterassignment\setshiftproofbelow\dimen0 }
\def\shiftproofbelowneg{\def\next{\multiply\dimen0 by-1 }%
\afterassignment\setshiftproofbelow\dimen0 }
\def\setshiftproofbelow{\next\proofbelowshift=\dimen0 }
\def\setproofrulebreadth{\proofrulebreadth}

\def\prooftree{
\ifnum	\lastpenalty=1
\then	\unpenalty
\else	\onleftofproofrulefalse
\fi
\ifonleftofproofrule
\else	\ifinsideprooftree
	\then	\hskip.5em plus1fil
	\fi
\fi
\bgroup
\setbox\proofbelow=\hbox{}\setbox\proofrulename=\hbox{}%
\let\justifies\proofover\let\leadsto\proofoverdots\let\Justifies\proofoverdbl
\let\using\proofusing\let\[\prooftree
\ifinsideprooftree\let\]\endprooftree\fi
\proofdotsfalse\doubleprooffalse
\let\thickness\setproofrulebreadth
\let\shiftright\shiftproofbelow \let\shift\shiftproofbelow
\let\shiftleft\shiftproofbelowneg
\let\ifwasinsideprooftree\ifinsideprooftree
\insideprooftreetrue
\setbox\proofabove=\hbox\bgroup$\displaystyle 
\let\wereinproofbit\prooftree
\shortenproofleft=0pt \shortenproofright=0pt \proofbelowshift=0pt
\onleftofproofruletrue\penalty1
}

\def\eproofbit{
\ifx	\wereinproofbit\prooftree
\then	\ifcase	\lastpenalty
	\then	\shortenproofright=0pt	
	\or	\unpenalty\hfil		
	\or	\unpenalty\unskip	
	\else	\shortenproofright=0pt	
	\fi
\fi
\global\dimen0=\shortenproofleft
\global\dimen1=\shortenproofright
\global\dimen2=\proofrulebreadth
\global\dimen3=\proofbelowshift
\global\dimen4=\proofdotseparation
\global\mscount=\proofdotnumber
$\egroup  
\shortenproofleft=\dimen0
\shortenproofright=\dimen1
\proofrulebreadth=\dimen2
\proofbelowshift=\dimen3
\proofdotseparation=\dimen4
\proofdotnumber=\mscount
}

\def\proofover{
\eproofbit 
\setbox\proofbelow=\hbox\bgroup 
\let\wereinproofbit\proofover
$\displaystyle
}%
\def\proofoverdbl{
\eproofbit 
\doubleprooftrue
\setbox\proofbelow=\hbox\bgroup 
\let\wereinproofbit\proofoverdbl
$\displaystyle
}%
\def\proofoverdots{
\eproofbit 
\proofdotstrue
\setbox\proofbelow=\hbox\bgroup 
\let\wereinproofbit\proofoverdots
$\displaystyle
}%
\def\proofusing{
\eproofbit 
\setbox\proofrulename=\hbox\bgroup 
\let\wereinproofbit\proofusing
\kern0.3em$
}

\def\endprooftree{
\eproofbit 
  \dimen5 =0pt
\dimen0=\wd\proofabove \advance\dimen0-\shortenproofleft
\advance\dimen0-\shortenproofright
\dimen1=.5\dimen0 \advance\dimen1-.5\wd\proofbelow
\dimen4=\dimen1
\advance\dimen1\proofbelowshift \advance\dimen4-\proofbelowshift
\ifdim	\dimen1<0pt
\then	\advance\shortenproofleft\dimen1
	\advance\dimen0-\dimen1
	\dimen1=0pt
	\ifdim  \shortenproofleft<0pt
        \then   \setbox\proofabove=\hbox{%
			\kern-\shortenproofleft\unhbox\proofabove}%
                \shortenproofleft=0pt
        \fi
\fi
\ifdim	\dimen4<0pt
\then	\advance\shortenproofright\dimen4
	\advance\dimen0-\dimen4
	\dimen4=0pt
\fi
\ifdim	\shortenproofright<\wd\proofrulename
\then	\shortenproofright=\wd\proofrulename
\fi
\dimen2=\shortenproofleft \advance\dimen2 by\dimen1
\dimen3=\shortenproofright\advance\dimen3 by\dimen4
\ifproofdots
\then
	\dimen6=\shortenproofleft \advance\dimen6 .5\dimen0
	\setbox1=\vbox to\proofdotseparation{\vss\hbox{$\cdot$}\vss}
	\setbox0=\hbox{%
		\kern\dimen6
		$\vcenter to\proofdotnumber\proofdotseparation
			{\leaders\box1\vfill}$%
		\unhbox\proofrulename}%
\else	\dimen6=\fontdimen22\the\textfont2 
	\dimen7=\dimen6
	\advance\dimen6by.5\proofrulebreadth
	\advance\dimen7by-.5\proofrulebreadth
	\setbox0=\hbox{%
		\kern\shortenproofleft
		\ifdoubleproof
		\then	\hbox to\dimen0{%
			$\mathsurround0pt\mathord=\mkern-6mu%
			\cleaders\hbox{$\mkern-2mu=\mkern-2mu$}\hfill
			\mkern-6mu\mathord=$}%
		\else	\vrule height\dimen6 depth-\dimen7 width\dimen0
		\fi
		\unhbox\proofrulename}%
	\ht0=\dimen6 \dp0=-\dimen7
\fi
\let\doll\relax
\ifwasinsideprooftree
\then	\let\VBOX\vbox
\else	\ifmmode\else$\let\doll=$\fi
	\let\VBOX\vcenter
\fi
\VBOX	{\baselineskip\proofrulebaseline \lineskip.2ex
	\expandafter\lineskiplimit\ifproofdots0ex\else-0.6ex\fi
	\hbox	spread\dimen5	{\hfi\unhbox\proofabove\hfi}%
	\hbox{\box0}%
	\hbox	{\kern\dimen2 \box\proofbelow}}\doll%
\global\dimen2=\dimen2
\global\dimen3=\dimen3
\egroup 
\ifonleftofproofrule
\then	\shortenproofleft=\dimen2
\fi
\shortenproofright=\dimen3
\onleftofproofrulefalse
\ifinsideprooftree
\then	\hskip.5em plus 1fil \penalty2
\fi
}

% ...

\mathcode`\<="4268  
\mathcode`\>="5269  

\mathchardef\gt="313E
\mathchardef\lt="313C

\newcommand{\NN}{\mathbb{N}}

\newcommand{\seqsym}{\vdash}

\newcommand{\seq}[2]{#1 \seqsym #2} 

\newcommand{\tjud}[3]{#1\ \rhd\ #2 \seqsym #3}
\newcommand{\pjud}[2]{#1 \seqsym #2}

\newcommand{\urule}[3]{%
  \prooftree #1 \justifies #2 \using #3 \endprooftree}
\newcommand{\brule}[4]{%
  \prooftree #1\ \ \ #2 \justifies #3 \using #4 \endprooftree}
\newcommand{\trule}[5]{%
  \prooftree #1\ \ \ #2 \ \ #3\justifies #4 \using #5 \endprooftree}

\newcommand{\imp}{\to} 

\newcommand{\clambda}{\overline{\lambda}}

\newcommand{\pl}{\textsf{\mdseries\upshape p}}
\newcommand{\tl}{\textsf{\mdseries\upshape t}}
\newcommand{\ql}{\textsf{\mdseries\upshape q}}
\newcommand{\el}{\textsf{\mdseries\upshape e}}
\newcommand{\pf}[1]{{#1^{\pl}}}
\newcommand{\tf}[1]{{#1^{\tl}}}
\newcommand{\qf}[1]{{#1^{\ql}}}
\newcommand{\ptq}{\mbox{\pl\tl\ql}}

\newcommand{\CT}{\textsf{KT}}
\newcommand{\LK}{\textsf{LK}}
\newcommand{\mT}{\textsf{mT}}

\newcommand{\ptqtr}[1]{\overline{#1}}
\newcommand{\cbn}[1]{\ptqtr{#1}^n}
\newcommand{\cbv}[1]{\ptqtr{#1}^v}
\newcommand{\auxptqtr}[1]{\underline{#1}}
\newcommand{\auxcbn}[1]{\auxptqtr{#1}_n}
\newcommand{\auxcbv}[1]{\auxptqtr{#1}_v}
\newcommand{\varptqtr}[1]{\overline{\overline{#1}}}
\newcommand{\varcbn}[1]{\varptqtr{#1}^n}
\newcommand{\varcbv}[1]{\varptqtr{#1}^v}
\newcommand{\plotr}[1]{\lceil{#1}\rceil}
\newcommand{\plocbn}[1]{\plotr{#1}^n}
\newcommand{\plocbv}[1]{\plotr{#1}^v}
\newcommand{\auxplotr}[1]{\lfloor{#1}\rfloor}
\newcommand{\auxplocbn}[1]{\auxplotr{#1}_n}
\newcommand{\auxplocbv}[1]{\auxplotr{#1}_v}
\newcommand{\rbk}[1]{\llbracket #1 \rrbracket}
\newcommand{\missn}[2]{|{#1}|_{#2}}

\newcommand{\contapp}[2]{#1 \circ #2}

\newcommand{\trar}{\rightsquigarrow}
\newcommand{\cbnar}{\stackrel{\mbox{\tiny CbN}}{\trar}}
\newcommand{\cbvar}{\stackrel{\mbox{\tiny CbV}}{\trar}}

\newcommand{\dummyvar}{\Box}
\newcommand{\dlambda}{\lambda_{\dummyvar}}
\newcommand{\dLambda}{\Lambda_{\dummyvar}}

\newcommand{\FV}[1]{\mathsf{FV}(#1)}

\newcommand{\id}{\mathsf{id}}

\newcommand{\redto}{\to}

\newcommand{\rredto}{\stackrel{*}{\redto}}

\newcommand{\cto}{\mapsto }
\newcommand{\cfrom}{\mapsfrom }

\newcommand{\mcto}{\stackrel{*}{\cto} }
\newcommand{\nto}{\cto_n }
\newcommand{\vto}{\cto_v }

\newcommand{\ev}{\Downarrow}
\newcommand{\cbvev}{\ev_v}
\newcommand{\cbnev}{\ev_n}

\newcommand{\lencntred}[1]{\ell({#1})}

\newdimen\netunit
\newlength\vt\newlength\hz
\newcommand{\lspine}[3]{%
  \save\POS
  @+, s0+<1em,1.5em>@+, s0+/u#3\vt/+/d1.5em/@+, s0+/l#2\hz/@+, 
  s0+/d#3\vt/@+, s0.{s0+/r#2\hz/}+R(.4)@+, s0+/u1.5em/@+,
  s6+<-1em,1.5em>@+,
  {s0 \ar@{-} 's1 's2 's3 's4 's5 's6 's7 s0},
  {s6.s4+C}.s1.s5="SPINE#1",
  s2.s3+C="SPINE#1-AUX",
  @-@-@-@-@-@-@-
  \restore\POS
  }

\newcommand{\rspine}[3]{%
  \save\POS
  @+, s0+<-1em,1.5em>@+, s0+/u#3\vt/+/d1.5em/@+, s0+/r#2\hz/@+, 
  s0+/d#3\vt/@+, s0.{s0+/l#2\hz/}+L(.4)@+, s0+/u1.5em/@+,
  s6+<1em,1.5em>@+,
  {s0 \ar@{-} 's1 's2 's3 's4 's5 's6 's7 s0},
  {s6.s4+C}.s1.s5="SPINE#1",
  s2.s3+C="SPINE#1-AUX",
  @-@-@-@-@-@-@-
  \restore\POS
  }

\title{%
  Continuations, proofs and tests\thanks{%
    Partially supported by the MIUR PRIN %
    ``Logical foundations of abstract programming languages''.} %
}

\author{%
  Stefano Guerrini\thanks{%
    \indent Dipartimento di Informatica, 
    Universit{\`a} Roma La Sapienza -
    Via Salaria, 113 -
    00198 Roma - Italy -
    email: \texttt{guerrini@dsi.uniroma1.it}}
  \and
  Andrea Masini\thanks{%
    \indent Dipartimento di Informatica, Universit{\`a} di Verona -
    Ca' Vignal 2, strada le Grazie, 15 - 37134 Verona - Italy -
    email: \texttt{andrea.masini@univr.it}}
  }

\begin{document}

\maketitle

\begin{abstract}
  \noindent
  Continuation Passing Style (CPS) is one of the most important issues
  in the field of functional programming languages, and the quest for
  a primitive notion of types for continuation is still open.\\
  Starting from the notion of ``test'' proposed by Girard, we develop a
  notion of test for intuitionistic logic. \\
  We give a complete deductive system for tests and we show that it is
  good to deal with ``continuations''. In particular, in the proposed
  system it is possible to work with Call by Value and Call by Name
  translations in a uniform way.
\end{abstract}


\section{Introduction}


Since the seminal works of Fischer~\cite{Fisch72} and
Plotkin~\cite{Plotk75} continuations have become central in the study
and implementation of functional languages. In particular, by defining
the Call-by-Name and Call-by-Value translations of $\lambda$-calculus,
Plotkin posed the basis of CPS (Continuation Passing Style)
transforms.

After these initial milestones, any overview of CPS, even if very
short, cannot avoid to mention the fundamental work that Felleisen,
Friedman, Kohlbecker, Duba, and Sabry carried out for type free
functional languages. Felleisen et al.~\cite{FellFried86} were the
first to axiomatize the so called control operators---\texttt{call/cc}
and other similar operators of lisp-like languages. Some years later,
Sabry and Felleisen~\cite{SabFell93} were the first to prove
completeness results for CPS transforms of type-free functional
languages.


While Felleisen and his collaborators were developing the theory of
CPS transforms, several researchers began to investigate if it was
possible to explain CPS by means of some known logic, in the style of
the well-known correspondence between intuitionistic logic and types
and computations of functional languages.

Griffin~\cite{Griff90} was the first one to give a partial answer to
this question by proposing classical logic as a type system for a
simplified version of Scheme. The main idea of Griffin was to use
Reductio ad Absurdum to explain Felleisen's control operator
$\mathcal{C}$.

Although the work of Griffin opened new perspectives in the use of
classical logic for the study of programming languages, it left
unanswered several questions. First of all, ``classical logic seemed
not to have a clear computational interpretation'' because of the lack
of confluence of its ``standard'' natural deduction formulation or, as
observed by Joyal in categorical terms, because of the collapse of
proofs in the standard categorical semantics of classical logic. But,
what does it happen if we change the rules of the game, namely the
``formulation of the logic''?

In~\cite{Par92,Par97}, Parigot showed that a drastically different
formulation of classical logic, the so-called $\lambda\mu$-calculus,
allows to give a computational meaning to the cut elimination
procedure---$\lambda\mu$-calculus enjoys the nice computational
properties of $\lambda$-calculus: strong normalization and confluence.

After the introduction of the $\lambda\mu$-calculus, several
researchers tried to show that it might have been a foundational
calculus for CPS (e.g., de~Groote~\cite{degr98}). Unfortunately, such
a research did not led to the expected results: it pointed out many
analogies between $\lambda\mu$-calculus and continuations, but, at the
same time, it showed that $\lambda\mu$-calculus fails to give a
precise definition of basic control operators. Indeed, it showed that
even if the $\mu$-reduction has a ``continuation flavor'', it is not
the right reduction for CPS transformed programs. In
spite of these negative results, $\lambda\mu$-calculus remains one of
the most important logical calculi for CPS.


In~\cite{HofStr97}, Hofmann and Streicher proposed a categorical
continuation model for a Call-by-Name version of
$\lambda\mu$-calculus. As already done by Griffin, Hofmann and
Streicher used classical logic, and in particular the Reduction ad
Absurdum principle, to define the meaning of CPS. Anyhow, while
Griffin used Reduction ad Absurdum to give a type assignments to CPS
terms, Hofmann and Streicher embedded Reduction ad Absurdum in the
construction of the semantic domains for the interpretation of CPS.

Subsequently, in~\cite{StrReus98}, Streicher and Reus extended the
ideas in~\cite{HofStr97} giving a categorical semantics of a
Call-by-Value $\lambda$-calculus equipped with the control operator
$\mathcal{C}$ of Felleisen.

Recently, a very interesting analysis of CPS in terms of proof/type
theoretical methods has been proposed by Curien and Herbelin:
in~\cite{CurHerb00}, they have shown that the most known CPS
translations may be obtained by means of a suitable translation
between $\lambda\mu$-calculus and a new formulation of
$\lambda$-calculus plus control operators. Another interesting
proof-theoretical contribution is the work of Ogata~\cite{Oga02}, who
related a Call-by-Value normalization of the $\lambda\mu$-calculus
with the cut-elimination of one of the logical systems proposed by
Danos, Joinet and Schellinx for the analysis/embedding of classical
logic trough/into linear logic.

The results of Hoffman, Streicher and Reus~\cite{HofStr97,StrReus98}
were the natural background for the introduction of Selinger's Control
Categories~\cite{Sel01}, perhaps, one the most important steps towards
a semantic/logical explanation of CPS. In fact, Control Categories
were the first model of $\lambda\mu$-calculus in which Call-by-Value
and Call-by-name have a uniform interpretation.

Few years after the work of Selinger, F\"uhrmann and
Thielecke~\cite{FuhThi04} presented a quite different approach to the
semantics of CPS---even if, restricted to the case of an idealized
Call-by-Value functional language. In particular, they proposed both a
type theoretical and a categorical semantics approach to CPS, and
studied in detail the CPS transforms.

\subsection{Interaction}

The problem of a satisfactory logical explanation of continuations
remains open.

Quite naturally, one may observe that any solution to such a problem
must base on a deep interaction between programs and
computations. Therefore, a good question is: in logic, is there any
explicit notion of ``interaction'' that could be used in CPS? The
positive answer, in our opinion, is in the proposal of Girard for an
``interactive approach to logic''.

The key point of Girard~\cite{Gir:Meaning1:99} is the idea that the
meaning of proofs does not reside in some external world called the
``semantics of the proofs''; the meaning of proofs is described by the
interaction between proofs and some dual objects that Girard name
\emph{tests}.

The proof/test duality introduced by Girard can be understood in terms
of a game between a player and an opponent. A \emph{proof} is a
sequence of arguments used by the player to assert that, moving from a
given set of premises, the ending formula (or sequent) holds. Then,
what is the dual of a proof? A \emph{test} is a sequence of
arguments used by the opponent to confute the provability of a formula
(or sequent). 

What does it happen if the player asserts that a formula is provable
while the opponent says that such a formula is not provable? If the
system is not trivial---and by the way, we are interested in such a
case only---someone is cheating and we need a way to validate the
arguments used in a proof/test. In this kind of game there is
no referee and we cannot resort to any external argument. So, the only
way that we have to discover who is cheating is by counterposing the
proof proposed by the player to the test proposed by the opponent. The
interaction between the two derivations (cut-elimination) will lead
to discover where the arguments of the player or of the opponent fail.

Another important issue is constructiveness: if we do not want to exit
outside our computational world, both proof and tests must be
constructive.

The BHK interpretation asserts that: 
\begin{quotation}
  \noindent
  A \emph{proof} $\pi$ of $A \imp B$ is a (constructive)
  transformation from a proof of $A$ to a proof of $B$.
\end{quotation}
In particular, if there is no proof of $A$, the transformation $\pi$
is the empty map, and we do not have any argument to refute it. On the
other hand, given a proof of $A$, the transformation $\pi$ leads to a
proof of $B$ that we can attempt to refute. Therefore, it is
quite natural to assert that: 
\begin{quotation}
  \noindent
  a \emph{test} $t$ of $A \imp B$ is a pair $(\pi,\tau)$ such that:
  \begin{enumerate}
  \item[(i)] $\pi$ is a proof of $A$ and 
  \item[(ii)] $\tau$ is a test of $B$.
  \end{enumerate}
\end{quotation}

Asking at the same time that:
\begin{quotation}
 a proof is a ``failure'' of a test and
 a test is a ``failure'' of a proof.
\end{quotation}
Such a notion of duality has been our starting point in the
development of a type system for continuations.

\subsection{Our proposal}

We propose a new calculus, the \ptq-calculus, characterized by the
relevant properties summarized below.
\begin{enumerate}
\item The \ptq-calculus bases on a \emph{general primitive notion of
    continuation/test}. That \emph{unique notion of continuation} is
  suitable to deal with both Call-by-Value and Call-by-Name languages.
\item The \ptq-calculus is equipped with a \emph{deterministic} one
  step lazy reduction relation: \emph{the calculus is, per se, neither
    Call-by-Value nor Call-by-Name}. A term is either in normal form
  or a redex.
\item Even if the \ptq-calculus is neither Call-by-Value nor
  Call-by-Name, \emph{it can code (in a sound and complete way)
    Call-by-Value and Call-by-Name $\lambda$-calculi}.
\end{enumerate}


\section{Proof theoretical motivations}

The technical details of the \ptq-calculus will be presented in
section~\ref{sec:ptq-calculus}. In this section, we shall give a
detailed and informal explanations of the proof theoretical
motivations that have led us to the calculus of continuations.

\subsection{The starting point: classical logic}

In the introduction, we have already seen that
\begin{itemize}
\item a test $t$ of $A \imp B$ is a pair $(\pi,\tau)$, where
  \begin{enumerate}
  \item $\pi$ is a proof of $A$ 
  \item[] and 
  \item $\tau$ is a test of $B$;
  \end{enumerate}
\item we have the following proof/test duality:
  \begin{itemize}
  \item a proof is a \textbf{failure} of a test;
  \item a test is a \textbf{failure} of a proof.
  \end{itemize}
\end{itemize}

Starting from these basic properties, the definition of test can be
extended in order to obtain a sound and complete proof system. The
system has two kind of formulas:
\begin{itemize}
\item \emph{proof formulas}, denoted by $\pf{A}$;
\item \emph{test formulas}, denoted by $\tf{A}$.
\end{itemize}

The judgments of the calculus are sequents of the form $\Gamma \vdash
\alpha$, where $\Gamma$ is a set of proof and test formulas, and
$\alpha$ is either empty, or a proof formula, or a test formula.

The following are the rules of the proof system---let us call it \CT.

$$\begin{array}{c@{\hspace{3ex}}c}
  \Gamma, \pf{A}\vdash \pf{A}
  &
  \Gamma, \tf{A}\vdash  \tf{A}
  \\[+2ex]
  \urule %
  { %
    {\Gamma}, \pf{A}, \tf{B}\vdash  %
  } %
  { %
    {\Gamma} \vdash \pf{A\to B} %
  } %
  {} %
  &
  \brule %
  { %
    {\Gamma}\vdash \pf{A}
  } %
  { %
    {\Gamma}\vdash \tf{B}
  } %
  { %
    {\Gamma}\vdash \tf{A\to B}
  } %
  { } %
  \\[+4ex]
  \urule %
  { %
    {\Gamma}, \tf{A}\vdash
  } %
  { %
    {\Gamma}\vdash \pf{A}
  } %
  { } %
  &
  \urule %
  { %
    {\Gamma},\pf{A}\vdash 
  } %
  { %
    {\Gamma}\vdash \tf{A}
  } %
  { } %
  \\[+4ex]
  \multicolumn{2}{c}{%
    \brule %
    { %
      {\Gamma} \vdash \tf{A}
    } %
    { %
      {\Gamma}\vdash \pf{A}
    } %
    { %
      {\Gamma} \vdash
    } %
    { } %
  }
\end{array}$$

It is quite easy to prove that \CT\ is a presentation of classical
logic.
 
\begin{proposition}
  The sequent $\Gamma\vdash \Delta$ is derivable in \LK\ iff
  the judgment $\pf{\Gamma},\tf{\Delta}\vdash$ is derivable in
  \CT.
\end{proposition}

As a direct consequence, \emph{$A$ is classically valid iff there
  exists a derivation of $\ \vdash \pf{A}$ in \CT}.

\subsection{Leaving the classical world \ldots}

It is immediate to observe that the proof-test duality is reminiscent
of the well-known de~Morgan duality: if we translate each $\pf{A}$ as
$A$ and each $\tf{A}$ as $\neg A$ in the above proposed deductive
system, we obtain a set of admissible rules for \LK. But, in
spite of such a connection, the proof-test duality does not introduce
any kind of classical principle, and in fact it will be used in an
\emph{intuitionistic} setting.

Looking carefully at the proposed system, it is possible to observe
that:
\begin{itemize}
\item a premise $\tf{A}$ morally corresponds to a conclusion $A$;
\item a conclusion $\tf{A}$ morally corresponds to a premise $A$;
\item a premise $\pf{A}$ directly corresponds to a premise $A$;
\item a conclusion $\pf{A}$ directly corresponds to a conclusion $A$.
\end{itemize}

As a matter of fact, it is possible to translate each judgment $G$ of
\CT\ in an ordinary sequent $S=(G)^+$ of \LK:
\begin{itemize}
\item  $(\pf{\Gamma},\tf{\Delta}\vdash)^+ = \Gamma\vdash\Delta$;
\item  $(\pf{\Gamma},\tf{\Delta}\vdash\pf{A})^+ = \Gamma\vdash\Delta, A$;
\item  $(\pf{\Gamma},\tf{\Delta}\vdash\tf{A})^+ = \Gamma, A\vdash\Delta$;
\end{itemize}

Such a translation, when applied to the rules of \CT, produces
the rules:
\begin{displaymath}
  \Gamma, {A}\vdash \Delta, {A} \qquad
  \urule %
  { %
    {\Gamma}, {A}\vdash \Delta, B  %
  } %
  { %
    {\Gamma} \vdash \Delta,  {A\to B} %
  } %
  {} %
  \qquad 
  \brule %
  { %
    {\Gamma}\vdash \Delta, {A}
  } %
  { %
    {\Gamma}, B\vdash \Delta
  } %
  { %
    {\Gamma},{A\to B}\vdash \Delta
  } %
  { } %
  \qquad   
  \brule %
  { %
    {\Gamma}, A \vdash \Delta
  } %
  { %
    {\Gamma}\vdash\Delta,  {A}
  } %
  { %
    {\Gamma} \vdash \Delta
  } %
  { } %
\end{displaymath}
which are the standard \LK\ rules of classical logic.
 
Now, let us consider \emph{minimal logic}, i.e., the system of types
for simply typed $\lambda$-calculus. We know that minimal logic is
obtained by means of a structural constraint: \emph{the sequents must
  have exactly one conclusion}.

If we want that the $(\ )^+$ translation produces minimal logic
sequents, we must constrain the structure of derivable judgments in
such a way that:
\begin{itemize}
\item[C1.] for each judgment $\Gamma\vdash\pf{A}$, the set $\Gamma$
  does not contain test formulas;
\item[C2.] for each judgment $\Gamma\vdash$, the set $\Gamma$ contains
  exactly one test formula;
\item[C3.] for each judgment $\Gamma\vdash\tf{A}$, the set $\Gamma$
  contains exactly one test formula.
\end{itemize}

The simpler way to obtain a deductive system such that all the
derivable judgments obey to the constraints (C1), (C2) and (C3) is to
impose a \emph{linear discipline for test formulas}, as in the 
deductive system below---let us call it \mT.

$$\begin{array}{c@{\hspace{3ex}}c}
  \pf{\Gamma}, \pf{A}\vdash \pf{A}
  &
  \pf{\Gamma}, \tf{A}\vdash  \tf{A}
  \\[+2ex]
  \urule %
  { %
    \pf{\Gamma}, \pf{A}, \tf{B}\vdash  %
  } %
  { %
    \pf{\Gamma} \vdash \pf{A\to B} %
  } %
  {} %
  &
  \brule %
  { %
    \pf{\Gamma}\vdash \pf{A}
  } %
  { %
    \pf{\Gamma},\tf{C}\vdash \tf{B}
  } %
  { %
    \pf{\Gamma}, \tf{C}\vdash \tf{A\to B}
  } %
  { } %
  \\[+4ex] 
  \urule %
  { %
    \pf{\Gamma}, \tf{A}\vdash
  } %
  { %
    \pf{\Gamma}\vdash \pf{A}
  } %
  { } %
  &
  \urule %
  { %
    \pf{\Gamma},\pf{A}, \tf{B}\vdash
  } %
  { %
    \pf{\Gamma}, \tf{B}\vdash \tf{A}
  } %
  { } %
  \\[+4ex]
  \multicolumn{2}{c}{%
    \brule %
    { %
      \pf{\Gamma} , \tf{B}\vdash \tf{A}
    } %
    { %
      \pf{\Gamma}\vdash \pf{A}
    } %
    { %
      \pf{\Gamma} ,\tf{B}\vdash
    } %
    { } %
  }
\end{array}$$

It is possible to prove that \mT\ is a presentation of minimal logic.

\begin{proposition}
  The sequent $\Gamma\vdash A$ is derivable in minimal logic iff the
  judgment $\pf{\Gamma}\vdash\pf{A}$ is derivable in \mT.
\end{proposition}

\subsubsection{... and approaching to continuations.}

In the perspective of the development of a type theory for
continuations, we think that the notion test described above is the
right one. Therefore, let us propose an extension of the standard
Curry-Howard isomorphism, by providing a correspondence between:
\begin{itemize}
\item \emph{deductions of $\pf{A}$ and programs of type $\pf{A}$};
\item \emph{deductions of $\tf{A}$ and continuations of type
    $\tf{A}$}.
\end{itemize}

As a first step, let us transform \mT\ into a type system.

$$\begin{array}{c@{\hspace{3ex}}c}
  \pf{\Gamma},x: \pf{A}\vdash x: \pf{A} &
  \pf{\Gamma}, k:\tf{A}\vdash  k:\tf{A}
  \\[+2ex]
  \urule %
  { %
    \pf{\Gamma}, x:\pf{A}, k:\tf{B}\vdash u %
  } %
  { %
    \pf{\Gamma}, \vdash \lambda<x^{\pf{A}},k^{\tf{B}}>.u:\pf{A\to B} %
  } %
  {\pf{\to}} %
  &
  \brule %
  { %
    \pf{\Gamma}\vdash p:\pf{A}
  } %
  { %
    \pf{\Gamma}, k:\tf{C}\vdash t:\tf{B}
  } %
  { %
    \pf{\Gamma}, k:\tf{C}\vdash <p,t>:\tf{A\to B}
  } %
  {\tf{\to} } %
  \\[+4ex]
  \urule %
  { %
    \pf{\Gamma}, k:\tf{A}\vdash u
  } %
  { %
    \pf{\Gamma}\vdash \lambda k^{\tf{A}}.u:\pf{A}
  } %
  { \pf{\lambda}} %
  &
  \urule %
  { %
    \pf{\Gamma},x:\pf{A}, k:\tf{B}\vdash u
  } %
  { %
    \pf{\Gamma}, k:\tf{B}\vdash \lambda x^{\pf{A}}.u:\tf{A}
  } %
  { \tf{\lambda}} %
  \\[+4ex]
  \multicolumn{2}{c}{%
    \brule %
    { %
      \pf{\Gamma} , k:\tf{B}\vdash t:\tf{A}
    } %
    { %
      \pf{\Gamma}\vdash p:\pf{A}
    } %
    { %
      \pf{\Gamma} , k:\tf{B}\vdash t\bullet p
    } %
    {@} %
  } 
\end{array}$$

The reduction rules for terms that naturally arise from the above
syntax are:
\begin{eqnarray*}
  <p,t>\bullet\lambda<x,k>.u & \to & u[p/x,t/k] \\
  \lambda x.u\bullet p  & \to & u[p/x] \\
  t\bullet\lambda k.u & \to & u[t/k]
\end{eqnarray*}

Unfortunately, such a system has the main defect of any naive term
system associated to classical logic: it is \emph{non-confluent}. In
fact,
$$
\xymatrix{%
  & *++{\lambda x.u' \bullet \lambda k.u''}
  \ar@{->}[dl] \ar@{->}[dr] & \\
  *++{u'[\lambda k.u''/x]} &*++{} & *++{u''[\lambda x.u'/k]}
}
$$
and there is no general way to close the diagram.

The non-confluence of the calculus cannot be solved by imposing a
fixed reduction strategy for $\lambda x.u' \bullet \lambda k.u$. There
is not a standard way to make a choice between $\lambda x.u' \bullet
\lambda k.u \to u'[\lambda k.u''/x]$ and $\lambda x.u' \bullet \lambda
k.u \to u''[\lambda x.u'/k]$. Each of the two possible choices implies
serious problem in normalization. It is exactly the problem of
reducing a cut between $\neg A$ and $A$ in classical logic.

Moreover, there is a ``programming language'' reason forcing to reject
the choice of fixed reduction strategy for $\lambda x.u' \bullet
\lambda k.u$.

By assuming that $\lambda x.u' \bullet \lambda k.u$ always reduces to
$u'[\lambda k.u''/x]$, we impose that \emph{continuation are
  ``functions''}; on the other hand, by assuming $\lambda x.u' \bullet
\lambda k.u$ always reduce to $u''[\lambda x.u'/k]$, we impose that
\emph{continuations are ``arguments''}. But, unfortunately, both that
choices do not agree with the continuation passing style translations
of Call-by-Value and Call-by-Name functional languages (e.g.,
see~\cite{Plotk75}). In other words, \emph{``we cannot statically
  determine whether a continuation is an argument or a function''}.

In order to solve the problem of composition between tests
(continuations) and proofs (programs) we propose:
\begin{enumerate}
\item a ``new class of types'' $\qf{A}$, where is any intuitionistic
  type, s.t.\ $\qf{A}$ is a subtype of $\pf{A}$;
\item two different ways for composing a program $p$ and a
  continuation $t$:
  \begin{itemize}
  \item a standard composition $pt$, in which the continuation is an
    argument;
  \item a dynamic composition $t;p$, where which term plays the role
    of the argument is not statically fixed (it could be either $t$ or
    $p$), depending on the shape of $t$ and $p$.
  \end{itemize}
\end{enumerate}


\section{The \ptq-calculus}
\label{sec:ptq-calculus}

The set of the \emph{type expressions} is given by the following
grammar:
$$\begin{array}{lcll}
  X   & ::= & X_1\ \mid\ \ldots\ \mid\ X_k & 
  \qquad \mbox{base types} \\
  A   & ::= & X\ \mid\ A \to A  &  \qquad \mbox{intuitionistic types}\\
  P   & ::= & \pf{A}  &  \qquad \mbox{proof types}\\
  T   & ::= & \tf{A}  &  \qquad \mbox{test types}\\
  Q   & ::= & \qf{A} 
  &  \qquad \mbox{\ql-proof types}\\
  W   & ::= & P\ \mid\ T\ \mid\ Q & \qquad \mbox{types}\\
\end{array}$$

The set of the \emph{\ptq-term expressions}, or \emph{\ptq-terms} for
short, is defined by the following grammar:
$$\begin{array}{lcll}
  x & ::= & x_0, x_1,\ldots & \qquad \mbox{\pl-variables}\\
  k & ::= & k_0, k_1,\ldots & \qquad \mbox{\tl-variables}\\
  p & ::= & x\ \mid\ \lambda<x,k>.u
             \ \mid\ \lambda k.u & \qquad \mbox{\pl-terms}\\
  t & ::= & *\ \mid\ k\ \mid\ <p,t>
             \ \mid\ \lambda x.u & \qquad\mbox{\tl-terms}\\
  q & ::= & \clambda k.u &\qquad\mbox{\ql-terms}\\
  u & ::= & t;p\mid qt & \qquad\mbox{\el-terms}
\end{array}$$

In order to simplify the treatment of substitution, we shall assume to
work modulo variable renaming, i.e., term-expressions are equivalence
classes modulo $\alpha$-conversion. Substitution up to
$\alpha$-equivalence is defined in the usual way.

\subsection{The type system}
\label{sec:type-system}

A \emph{type environment} $\Xi$ is either a set $\Gamma$ or a pair of
sets $\Gamma \rhd \delta$, where $\Gamma$ is a (possibly empty) set
$x_1:P_1, \ldots, x_n:P_n$ of typed \pl-variables, such that all the
variables $x_i$ are distinct, and $\delta$ is either a singleton
$k:\tf{A}$ (a typed \tl-variable) or a singleton $*:\tf{A}$ (a typed
constant).

A \emph{judgment} is an expression $\Xi \vdash \xi$, where $\Xi$
is a type environment, $\xi$ is a typed \ptq-term expression, and
all the free variables in $\xi$ occur in $\Gamma$.

The set of the \emph{well-typed \ptq-terms} and the set of the
\emph{well-typed judgments} are defined by the type system in
Figure~\ref{fig:ptq-type-system}, where $A$ and $B$ are metavariables
ranging over intuitionistic types. (In the following, when not
otherwise specified, the metavariables $A,B,C,...$ will range over
intuitionistic types.)

\begin{figure}[htbp]
  \centering
$\begin{array}{c@{\qquad}c}
    \pjud{{\Gamma}, x:\pf{A}}{x:\pf{A}} &
    \tjud{{\Gamma}}{k:\tf{A}}{k:\tf{A}}
    \\[2ex]
    & \tjud{{\Gamma}}{*:\tf{A}}{*:\tf{A}}
    \\[2ex]
    \urule %
    { %
      \tjud{{\Gamma}, x:\pf{A}}{k:\tf{B}}{u} %
    } %
    { %
      \pjud{{\Gamma}}{\lambda<x,k>.u:\pf{A\to B}} %
    } %
    {\pf{\to}} %
    &
   \brule %
   { %
     \pjud{{\Gamma}}{}{p:\pf{A}}
   } %
   { %
     \tjud{{\Gamma}}{\delta}{t:\tf{B}}
   } %
   { %
     \tjud{{\Gamma}}{\delta}{<p,t>:\tf{A\to B}}
   } %
   {\tf{\to} } %
   \\[4ex]
   \urule %
   { %
     \tjud{\Gamma}{k:\tf{A}}{u}
   } %
   { %
     \pjud{\Gamma}{\lambda k}.u:\pf{A}
   } %
   { \pf{\lambda}} %
   &
   \urule %
   { %
     \tjud{\Gamma, x:\pf{A}}{\delta}{u}
   } %
   { %
     \tjud{\Gamma}{\delta}{\lambda x}.u:\tf{A}
   } %
   { \tf{\lambda}} %
   \\[4ex]
\urule %
   { %
     \tjud{\Gamma}{k:\tf{A}}{u}
   } %
   { %
     \pjud{\Gamma}{\clambda k.u:\qf{A}}
   } %
   { \qf{\clambda}} %
&
\\[4ex]
\multicolumn{2}{c}{%
\brule %
   { %
     \pjud{\Gamma}{p:\pf{A}}
   } %
   { %
     \tjud{\Gamma}{\delta}{t:\tf{A}}
   } %
   { %
     \tjud{\Gamma}{\delta}{t;p}
   } %
   {\pf{@}} %
}
\\[4ex]
\multicolumn{2}{c}{%
\brule %
   { %
     \pjud{\Gamma}{q:\qf{A}} 
   } %
   { %
     \tjud{\Gamma}{\delta}{t:\tf{A}}
   } %
   { %
     \tjud{\Gamma}{\delta}{qt}
   } %
   {\qf{@}} %
}
\end{array}$  
  \caption{\ptq-type system}
  \label{fig:ptq-type-system}
\end{figure}

By inspection of the type system in Figure~\ref{fig:ptq-type-system},
we see that:
\begin{itemize}
\item in well-typed \ptq-terms, \tl-variables are linear;
\item in a well-typed \pl-term/\ql-term there are no free occurrences
  of \tl-variables and no occurrences of the constant $*$;
\item in a well-typed \tl-term/\el-term there is one free occurrence
  of a \tl-variable or, alternatively, one occurrence of the constant
  $*$, that in any case cannot be enclosed by a \tl-variable binder.
\end{itemize}

Summing up, in order to construct well-typed terms, we suffice one
name for \tl-variables. Therefore, in the following, we shall
assume that all the occurrences of \tl-variables have name $k$.

\begin{fact}
  The \ptq-type system has the substitution property.
  \begin{enumerate}
  \item For any well-typed $\tjud{\Gamma}{\xi}{t':\tf{A}}$, we have that
    \begin{enumerate}
    \item for every well-typed
      $\tjud{\Gamma,\Delta}{k:\tf{A}}{t:\tf{B}}$ or
      $\tjud{\Gamma,\Delta}{*:\tf{A}}{t:\tf{B}}$,
      the corresponding
      $\tjud{\Gamma,\Delta}{\xi}{t[t'/k]:\tf{B}}$ or
      $\tjud{\Gamma,\Delta}{\xi}{t[t'/*]:\tf{B}}$ is well-typed;
    \item for every well-typed $\tjud{\Gamma,\Delta}{k:\tf{A}}{u}$ or
      $\tjud{\Gamma,\Delta}{*:\tf{A}}{u}$,
      the corresponding
      $\tjud{\Gamma,\Delta}{\xi}{u[t'/k]}$ or
      $\tjud{\Gamma,\Delta}{\xi}{u[t'/*]:\tf{B}}$ is well-typed.
    \end{enumerate}
  \item For any well-typed $\pjud{\Gamma}{p':\pf{A}}$, we have that
    \begin{enumerate}
    \item for every well-typed
      $\tjud{\Gamma,\Delta,x:\pf{A}}{\xi}{t:\tf{B}}$, then
      $\tjud{\Gamma,\Delta}{\xi}{t[p'/x]:\tf{B}}$ is well-typed;
    \item for every well-typed $\pjud{\Gamma,\Delta,x:\pf{A}}{u}$ or
      $\pjud{\Gamma,\Delta,x:\pf{A}}{p:\pf{B}}$ or
      $\pjud{\Gamma,\Delta,x:\pf{A}}{q:\qf{B}}$,
      the corresponding
      $\pjud{\Gamma,\Delta}{u[p'/x]}$ or
      $\pjud{\Gamma,\Delta}{p[p'/x]:\pf{B}}$ or
      $\pjud{\Gamma,\Delta}{q[p'/x]:\qf{B}}$ 
      is well-typed.
    \end{enumerate}
  \end{enumerate}
\end{fact}

A term is \tl-closed when it does not contain free occurrences of
\tl-variables. We have already seen that every well-typed
\pl-term/\ql-term is \tl-closed and that every \tl-closed well-typed
\tl-term/\el-term contains either a free occurrence of the
\tl-variable $k$ or an occurrence of the constant $*$ outside the
scope of any \tl-binder. Thus, every \tl-closed well-typed \tl-term
$t_*$ or \el-term $u_*$ can be obtained by replacing $*$ for the free
\tl-variable $k$ in a well-typed \tl-term $t$ or \el-term $u$ that
does not contain any occurrence of the constant $*$, namely
$$\begin{array}{lcl@{\qquad\qquad\mbox{ with }\qquad\qquad}lcl}
  t_* & = & t[*/k] & t_*[k/*] & = & t \\ 
  u_* & = & u[*/k] & u_*[k/*] & = & u
\end{array}$$
respectively.

In the following, we shall only consider well-typed \ptq-terms and
well-typed judgments; therefore, we shall omit to specify that a term
or judgment is well-typed. When not otherwise specified, we
shall always denote \tl-closed \tl-terms or \el-terms with a $*$
subscript; moreover, given a \tl-closed \tl-term $t_*$ and a \tl-closed
\el-term $u_*$, the terms $t$ and $u$ are the terms such that $t_* =
t[*/k]$ and $u_* = u[*/k]$.

The $*$-composition of \tl-closed \tl-terms is defined by
$$\contapp{t_*}{t_*'}=t_*'[t_*/*]=t'[t_*/k]$$
It is readily seen that the $*$-composition is associative and that
$*$ is its neutral element.
Such a composition is extended to
$$\contapp{t_*}{u_*}=u_*[t_*/*]=u[t_*/k]$$
which corresponds to $\contapp{t_*}{(t_*';p)}=(\contapp{t_*}{t_*'});p$
and $\contapp{t_*}{qt_*'}=q(\contapp{t_*}{t_*'})$.



\subsection{Computations}

According to a standard \emph{lazy} approach, the reduction rules that
we shall define \emph{do not reduce inside the scope of a $\lambda$ or
  $\clambda$ binder and inside a pair}. Since
\begin{itemize}
\item a \emph{\pl-term} is either a variable or begins with a
  $\lambda$,
\item a \emph{\tl-term} is either a variable or begins with a
  $\lambda$ or it is a pair,
\item a \emph{\ql-term} begins with a $\clambda$,
\end{itemize}
\tl-terms, \pl-terms and \ql-terms are \emph{irreducible}.

Now, let us observe that every \el-term $u$ is the composition of
irreducible terms, namely $u=t;p$ or $u=qt$, where $t,p,q$ are
irreducible. In the \ptq-calculus, we do not have any kind of
evaluation context, nor any notion of reduction strategy: a term (a
\el-term) contains at most one redex and either it is in normal form
or it is the redex to be reduced. This is a main difference w.r.t.\
standard $\lambda$-calculi, where we may choose the order of
evaluation by fixing a reduction strategy, for instance, (lazy)
call-by-value or call-by-name. Moreover, even when we choose a
reduction strategy, the $\beta$-redex $R$ that we have to reduce may
be deeply nested into the term $T$, that is $T=\mathcal{C}[R]$, where
$\mathcal{C}[]$ is the evaluation context determined by the reduction
strategy.

The fact that in the \ptq-calculus \el-terms only can be redexes
corresponds to the intended interpretation that proofs are programs
and tests are continuations. In particular, in order to start the
execution of a program we need to pass a continuation to it, that is
we have to compose the \pl-term corresponding to the program that we
want to execute with the \tl-term corresponding to the continuation
that we want to pass to it.

In a \pl-composition $u=t;p$, the execution is controlled by the shape
of the \tl-term $t$ (the continuation), namely
\begin{enumerate}
\item when $t$ is a constant or a pair, the control passes to the
  \pl-term $p$. In particular, 
  \begin{enumerate}
  \item if $p=\lambda k.u$, then $p$ corresponds to a suspended
    execution that is waiting for a continuation to
    put in the place of the parameter $k$. When $p$ is applied to the
    \tl-term $t$, the variable $k$ is replaced with $t$ in the body of
    $p$ and the execution resumes;
  \item if $p=\lambda<x,k>.u$, then $p$ corresponds to a suspended
    execution that is waiting for a program to put in the place of the
    parameter $x$ and a continuation to put in the place of the
    parameter $k$. Then the term $t;p$ reduces only when $t=<p',t'>$
    is a pair; in that case, the variables $x$ and $k$ in $u$ are
    replaced by the program $p'$ and the continuation $t'$,
    respectively, and the execution resumes;
  \end{enumerate}
\item when $t$ is a $\lambda$-abstraction, the continuation
  corresponding to $t=\lambda x.u$ can be interpreted as a suspended
  execution waiting for the actual value of a parameter
  $x$. Therefore, after replacing the program $p$ for $x$ in $u$, the
  execution resumes.
\end{enumerate}

Summing up, we have the reduction rules
\begin{eqnarray*}
  *;\lambda k. u & \redto & u_*\\
  <p,t>;\lambda k. u & \redto & u[<p,t>/k]\\
  <p,t>;\lambda <x, k>. u & \redto & u[p/x,t/k] \\
  \lambda x. u; p & \redto & u[p/x]
\end{eqnarray*}

Let us remark that, when $t$ and $p$ are both $\lambda$-abstraction,
$t=\lambda x.u$ is a $\lambda$-abstraction, if $p=\lambda k. u'$ is a
$\lambda$-abstraction too, we might try to reduce $t;p=\lambda
x.u;\lambda k. u'$ by replacing $t$ for the variable $k$ in
$u'$. Unfortunately, such a reduction rule would lead to the critical
pair
$$u'[\lambda k.u/x] \quad \cfrom \quad
\lambda x. u'; \lambda k.u \quad \redto \quad 
u[\lambda x. u'/k]
$$

In order to avoid critical pairs, we assume that, when $t$ is a
$\lambda$-abstraction, $t$ is the function to be applied on the
argument $p$, independently from the shape of $p$.  In other words, we
have a dynamic reduction strategy, corresponding to a sequential
policy of the kind ``first fit'', which summarizes in the following
rule:

\begin{itemize}
\item[] reducing a \tl-application $t;p$
  \begin{enumerate}
  \item analyze first $t$ and then $p$;
  \item contract the application by assuming that the first term that is
    ``usable'' as a function receives the other term as an argument.
  \end{enumerate}
\end{itemize}

The previous reduction rule does not suffice for our purposes. In some
cases, namely for the encoding of call-by-value, we also need the rule
that takes the \tl-term $t$ as an argument of some kind of function
constructed by abstracting the only free \tl-variable $k$ in the
\el-term $u$, even when $t=\lambda x. u$ is a $\lambda$-abstraction at
its turn. For this reason, in the type system, we have an abstraction
$\clambda k.u: \qf{A}$ that, if $k:\tf{A}$, construct a term
$q:\qf{A}$, whose type is not a test type, and another application $q
t$, for which we have the only reduction rule
$$
\clambda x. u; t \quad \redto \quad u[t/k]
$$ 
Let us remark that, since any \ql-term is a $\clambda$-abstraction, a
term $qt$ is always a redex.

In the previous analysis we have omitted the case in which $t$ is a
variable. The reason is that, if we take $k;\lambda k. u \redto u$, we
do not have $(k;\lambda k.u)[\lambda x.u'/k] \redto u[\lambda x.u'/k]$,
but $(k;\lambda k.u)[\lambda x.u'/k] \redto u'[\lambda k.u/x]$. Because
of this, we prefer to assume that the term $k;p$ is always
irreducible.  This also explains the role of the constant $*$. Since,
such a constant cannot be bound, it cannot be interpreted as a
placeholder for an arbitrary \tl-term and there is no problem in
reducing the \el-term $*;\lambda k.u$ to $u_*$. Anyhow, we remark that
$*;\lambda<x,k>.u$ is irreducible.

In order to complete the explanation of the role of the constant $*$,
let us recall that we want to interpret a \pl-term $p$ as the
translation of a program. Since \el-terms only are reducible, in order
to start the computation of $p$ we have to composite it with some
\tl-term, that is to pass some continuation to the program. The
simplest choice is to compose $p$ with the constant $*$.
Correspondingly,
\begin{quotation}
  \noindent
  the constant $*$ plays the role of the initial continuation of the
  system: \emph{the continuation that the ``system'' passes to the
    compiled code in order to start the computation}.
\end{quotation}

This assumption is fully justified by the fact that, following a
Continuation Passing Style, we shall compile any $\lambda$-term into
\tl-closed \pl-term. Because of this, we can also
\begin{quotation}
  \noindent
  restrict the reduction rules to \tl-closed terms.
\end{quotation}

The complete set of the rules of the calculus are given in
Figure~\ref{fig:red-rules}.

\begin{figure}[htbp]
  \centering
  \begin{eqnarray*}
    *;\lambda k. u & \redto & u_*\\
    <p,t_*>;\lambda k. u & \redto & u[<p,t_*>/k]\\
    <p,t_*>;\lambda <x, k>. u & \redto & u[p/x,t_*/k] \\
    \lambda x. u_*; p & \redto & u_*[p/x] \\
    (\clambda k. u) t_* & \redto & u[t_*/k]
  \end{eqnarray*}
  \caption{The reduction rules of the \ptq-calculus 
    (restricted to \tl-closed terms)}
  \label{fig:red-rules}
\end{figure}

As usual, we shall denote by $\rredto$ the transitive and reflexive
closure of $\redto$.

One of the standard interpretation of a continuation is as the rest of
computation: the continuation passed to a program specifies how the
computation must continue after the completion of the program. (For a
comparison of this interpretation with the interpretation that thinks
at a continuation as an evaluation context,
see~\cite{Dan:EvalContContRestComp:04}.) Accordingly, we expect that
the reduction of $t;p$ starts by the reduction that mimic the
execution of the program corresponding to $p$ and that, only after the
completion of that program, the execution is resumed by the
continuation. In practice, we expect that $t_*;p \rredto
\contapp{t_*}{u_*} = u[t/k]$, whenever $*;p \rredto u_*$. However, it
is readily seen that this cannot hold if $t_*=\lambda x.u_*'$ and
$p=\lambda k.u_*''$.

Lemma~\ref{lem:red-subst} gives the exact condition under which we may
get the expected replacement property: either $t_*$ is not a
$\lambda$-abstraction or, when this is not the case, during the
reduction of $*;p$ we never apply the rule that reduces a term with
the shape $*;\lambda k.u_*''$. Let us remark that this condition hold
for the translations of $\lambda$-terms that we shall give in the
paper.

\begin{lemma}
  \label{lem:red-subst}
  Let $u_*$ be an \el-term s.t.\ $u_* \rredto u_*'$. Given a \tl-term
  $t_*$,
  \begin{enumerate}
  \item if $t_*$ is not an abstraction $\lambda x.u_*''$, or
  \item $u_* \rredto u_*'$ without reducing any redex with the shape
    $*;\lambda k.u_*''$,
  \end{enumerate}
  then $\contapp{t_*}{u_*}\rredto\contapp{t_*}{u_*'}$, for every
  closed \tl-term $t_*$.
\end{lemma}
\begin{proof}
  By inspection of the reduction rules, we see that, since we cannot
  have $u_*=\lambda x.u_*'';\lambda k. u_*'$, the statement holds for
  a one-step reduction. Then, by induction on the length of the
  reduction, we conclude.
\end{proof}

\subsection{Readback}

The \ptq-calculus can be translated into the $\lambda$-calculus by a
map that associates to each term of the \ptq-calculus a typed
$\lambda$-term. If we forget the types, every \ptq-term is mapped into
a $\lambda$-term that contains the same \pl-variables of the \ptq-term
and one special constant $\dummyvar$, named \emph{hole}, that plays
the role of the free \tl-variable in the \ptq-term.

A $\dlambda$-term is a $\lambda$-term that may contain occurrences of
the \emph{hole} and whose variables range over the set of the
\pl-variables. If $M$ and $N$ are $\dlambda$-terms, the
\emph{hole composition} is defined by
$$\contapp{M}{N} = M[N/\dummyvar]$$
The hole composition is associative and $\dummyvar$ is its neutral
element, for $\contapp{M}{\dummyvar} = M = \contapp{\dummyvar}{M}$.

The (untyped) \emph{readback} map $\rbk{\cdot}$ is defined in
Figure~\ref{fig:rb-map}. Every \tl-term is mapped into a corresponding
$\dlambda$-term that may contain holes, while the other kind of terms
are mapped into $\dlambda$-terms that does not contain holes.

\begin{figure}[htbp]
  \centering
$$\begin{array}{lcl@{\qquad\qquad}lcl}
  \rbk{*} & = & \dummyvar &
  \rbk{x} & = & x \\[1.1ex] 
  \rbk{<p,t_*>} & = & \contapp{\rbk{t_*}}{\dummyvar \rbk{p}} &
  \rbk{\lambda<x,k>.u} & = &\lambda x.\rbk{u_*} \\[1.1ex]
  \rbk{\lambda x.u_*} & = & \rbk{u_*}[\dummyvar/x] &
  \rbk{\lambda k.u} & = & \rbk{u_*} \\[1.5ex]
\end{array}
$$
$$\begin{array}{lcl}
  \rbk{\clambda k.u} & = & \rbk{u_*}
  \\[1.1ex]
  \rbk{t_*;p}  & = & \contapp{\rbk{t_*}}{\rbk{p}}
  \\[1.1ex]
  \rbk{qt_*} & = & \contapp{\rbk{t_*}}{\rbk{q}}
\end{array}$$
  \caption{The readback map}
  \label{fig:rb-map}
\end{figure}

The readback map naturally extends to judgments. The typed
$\dlambda$-terms obtained by the translation are terms of a typed
$\lambda$-calculus $\dLambda$ in which:
\begin{itemize}
\item the set of the base types is the same of the \ptq-calculus;
\item the set of the variables is the set of the \pl-variables of the
  \ptq-calculus;
\item for each type $A$, there is a constant $\dummyvar^A:A$, the
  \emph{hole} of type $A$;
\item the type assignment rules are those of the typed
  $\lambda$-calculus, with the restriction that
\item a term of $\dLambda$ cannot contain occurrences of holes with
  different types (in any case, a term of $\dLambda$ may contain more
  than one hole of the same type);
\item the reduction rule is the standard $\beta$-reduction.
\end{itemize}

Let $\Gamma = x_1:{A_1},\ldots,x_n :{A_n}$ be a set of type
assignments for variables. We shall denote by $\pf{\Gamma} =
x_1:\pf{A_1},\ldots,x_n:\pf{A_n}$ the corresponding type assignment for
\pl-variables. The readback of typing judgments is defined by:
$$\begin{array}{lcl}
  \rbk{\tjud{\pf{\Gamma}}{k:\tf{A}}{t:\tf{B}}} & 
  = &\Gamma, \dummyvar: B \vdash \rbk{t_*}:A
  \\[1.1ex]
  \rbk{\tjud{\pf{\Gamma}}{*:\tf{A}}{t_*:\tf{B}}} &
  = &\Gamma,\dummyvar:B \vdash \rbk{t_*}:A
  \\[1.1ex]
  \rbk{\tjud{\pf{\Gamma}}{\phantom{}}{p:\pf{A}}} &
  = &\Gamma\vdash\rbk{p}:A 
  \\[1.1ex]
  \rbk{\tjud{\pf{\Gamma}}{\phantom{}}{q:\qf{A}}} &
  = &\Gamma\vdash\rbk{q}:A 
  \\[1.1ex]
  \rbk{\tjud{\pf{\Gamma}}{k:\tf{A}}{u}} &
  = &\Gamma\vdash\rbk{u_*}:A
  \\[1.1ex]
  \rbk{\tjud{\pf{\Gamma}}{*:\tf{A}}{u_*}} &
  = &\Gamma\vdash\rbk{u_*}:A
  \\[1.1ex]
  \end{array}
$$




\begin{remark}
  Typed $\dlambda$-terms with holes are a sort of typed
  \emph{contexts}. But, differently from the standard definition of
  contexts, $\dLambda$-terms are equivalence classes modulo variable
  renaming ($\alpha$-rule), for hole instantiation is not variable
  capturing. In a standard context, if a hole is in the scope of a
  $\lambda$-abstraction binding the variable $x$, the free occurrences
  of the variable $x$ in a term $M$ will be bound by the
  $\lambda$-abstraction when $M$ is put into the hole. In a
  $\dLambda$-term, the hole composition $\contapp{N}{M} =
  N[M/\dummyvar]$ is defined by means of the standard variable
  substitution; therefore, the variable $x$ bound in $N$ must be
  renamed and the free occurrences of $x$ in $M$ remain free in
  $\contapp{N}{M}$.
\end{remark}

\begin{proposition}
  Let the judgment $\tjud{\pf{\Gamma}}{\delta}{\xi}$ be derivable in the
  \ptq-calculus. The judgment $\rbk{\tjud{\pf{\Gamma}}{\delta}{\xi}}$ is
  derivable in $\dLambda$.
\end{proposition}
\begin{proof}
  By induction on the derivation of $\tjud{\Gamma^p}{\delta}{\xi}$.
\end{proof}

W.r.t.\ the readback, the $*$ plays the role of a neutral element,
namely $\rbk{*;p}=\rbk{p}$ and $\rbk{q*}=\rbk{q}$. Indeed, the
readback transformation maps the $*$-composition into the hole
composition. (Let us recall that the $*$-composition on \tl-terms is
defined by $\contapp{t_*}{t_*'} = t'[t_*/k]$.)

\begin{lemma}\label{lem:rb-composition}
  \mbox{}
  \begin{enumerate}
  \item $\rbk{\contapp{t_*}{t_*'}} = \contapp{\rbk{t_*}}{\rbk{t_*'}}$
  \item $\rbk{\contapp{t_*}{u_*}} = \contapp{\rbk{t_*}}{\rbk{u_*}}$
  \end{enumerate}
\end{lemma}
\begin{proof}
  By structural induction on $t'$ and $u$.
  \begin{enumerate}
  \item Let $A=\rbk{\contapp{t_*}{t_*'}}$ and
    $B=\contapp{\rbk{t_*}}{\rbk{t_*'}}$.
    \begin{enumerate}
    \item If $t' = k$, then $B = \contapp{\rbk{t_*}}{\rbk{*}} =
      \contapp{\rbk{t_*}}{\dummyvar} = \rbk{t_*} =
      \contapp{\rbk{t_*}{*}} = A$.
    \item If $t' = <t'',p>$, then %
      $A = \rbk{<\contapp{t_*}{t_*''},p>} =$ %
      (by the definition of readback) %
      $\contapp{\rbk{\contapp{t_*}{t_*''}}}{\dummyvar\rbk{p}} =$ %
      (by the induction hypothesis) %
      $\contapp{(\contapp{\rbk{t_*}}{\rbk{t_*''}})}{\dummyvar\rbk{p}} =
      \contapp{\rbk{t_*}}{(\contapp{\rbk{t_*''}}{\dummyvar\rbk{p}})} =$ %
      (by the definition of readback) %
      $\contapp{\rbk{t_*}}{\rbk{<t_*'',p>}} = B$. %
    \item If $t' = \lambda x.u$ (with $x\not\in\FV{t}$), then %
      $A = \rbk{\contapp{t_*}{\lambda x.u_*}} = %
      \rbk{\lambda x.(\contapp{t_*}{u_*})} = $ %
      (by the definition of readback) %
      $\rbk{\contapp{t_*}{u_*}}[\dummyvar/x] =$ %
      (by the induction hypothesis) %
      $(\contapp{\rbk{t_*}}{\rbk{u_*}})[\dummyvar/x] =$ %
      (by $x\not\in\FV{t}$) %
      $\contapp{\rbk{t_*}}{(\rbk{u_*}[\dummyvar/x])} =$ %
      (by the definition of readback) %
      $\contapp{\rbk{t_*}}{\rbk{\lambda x.u_*}} = B$.
    \end{enumerate}
  \item Let $A = \rbk{\contapp{t_*}{u_*}}$ and $B =
    \contapp{\rbk{t_*}}{\rbk{u_*}}$.
    \begin{enumerate}  
    \item If $u = t';p$, then %
      $A = \rbk{(\contapp{t}{t_*'});p} =$ %
      (by the definition of readback) %
      $\contapp{\rbk{\contapp{t_*}{t_*'}}}{\rbk{p}}=$ %
      (by the induction hypothesis) %
      $\contapp{(\contapp{\rbk{t_*}}{\rbk{t_*'}})}{\rbk{p}}=
      \contapp{\rbk{t_*}}{(\contapp{\rbk{t_*'}}{\rbk{p}})}=$ %
      (by the definition of readback) %
      $\contapp{\rbk{t_*}}{\rbk{t_*';p}} = B$.
    \item If $u = qt'$, then %
      $A = \rbk{q(\contapp{t}{t_*'})} =$ %
      (by the definition of readback) %
      $\contapp{\rbk{\contapp{t_*}{t_*'}}}{\rbk{q}}=$ %
      (by the induction hypothesis) %
      $\contapp{(\contapp{\rbk{t_*}}{\rbk{t_*'}})}{\rbk{q}}=
      \contapp{\rbk{t_*}}{(\contapp{\rbk{t_*'}}{\rbk{q}})}=$ %
      (by the definition of readback) %
      $\contapp{\rbk{t_*}}{\rbk{qt_*'}} = B$.
    \end{enumerate}
  \end{enumerate}
\end{proof}

Let us define $\contapp{t_*}{(t_*';p)}=(\contapp{t_*}{t_*'});p$ and
$\contapp{t_*}{(qt_*')}=q(\contapp{t_*}{t_*'})$.

\begin{corollary}\label{cor:rb-eval}
  $\rbk{\contapp{t_*}{u}} = \contapp{\rbk{t_*}}{\rbk{u_*}}$
\end{corollary}

Proposition~\ref{prp:rb-soundness} proves that the readback is sound
w.r.t.\ $beta$-reduction. In order to prove that proposition, we have
to show (Lemma~\ref{lem:rb-subst}) that the readback is sound w.r.t.\
\pl-variable substitution.

\begin{lemma}\label{lem:rb-subst}
  Let $a$ be a \tl-closed \tl-term or a \pl-term or a
  \ql-term or a \tl-closed \el-term. Then
  $$\rbk{a[p/x]} = \rbk{a}[\rbk{p}/x]$$
\end{lemma}
\begin{proof}
  By induction on the structure of $a$.
\end{proof}

\begin{proposition}\label{prp:rb-soundness}
  If $u \redto u'$, then $\rbk{u} \rredto \rbk{u'}$. Moreover,
  \begin{enumerate}
  \item if $u=<p,t_*>;\lambda <x, k>. u'' \redto u''[p/x,t_*/k] = u'$, then 
    $\rbk{u} = \redto \rbk{u'}$;
  \item otherwise, $\rbk{u} = \rbk{u'}$.
  \end{enumerate}
\end{proposition}
\begin{proof}
  When $u=<p,t_*>;\lambda <x, k>. u'' \redto u''[p/x,t_*/k]$, we have
  that
  \begin{multline*}
    \rbk{<p,t_*>;\lambda <x, k>. u''} \\ %
    = \contapp{\rbk{<p,t_*>}}{\rbk{\lambda <x,k>.u''}} %
    = \contapp{\rbk{<p,t_*>}}{\lambda x.\rbk{u_*''}}  %
    = \contapp{\rbk{t_*}}{(\lambda x.\rbk{u_*''})\rbk{p}} \\ %
    \redto \contapp{\rbk{t_*}}{\rbk{u_*''}[\rbk{p}/x]} %
    = \mbox{(by Lemma~\ref{lem:rb-subst}) } %
    \contapp{\rbk{t_*}}{\rbk{u_*''[p/x]}} \\  %
    = \rbk{\contapp{t_*}{u_*''[p/x]}} %
    = \rbk{u''[p/x,t_*/k]}
  \end{multline*}
  For the other reduction rules, we have instead
  \begin{itemize}
  \item $\rbk{*;\lambda k. u''} = \rbk{u_*''}$
  \item $\rbk{<p,t_*>;\lambda k. u''} %
    = \contapp{\rbk{<p,t_*>}}{\rbk{u_*''}} %
    = \rbk{\contapp{<p,t_*>}{u_*''}} %
    = \rbk{u''[<p,t*>/k]}$
  \item $\rbk{\lambda x.u_*'';p} %
    = \rbk{u_*''}[\rbk{p}/x] %
    = \rbk{u''[p/x]}$
  \item $\rbk{(\clambda k.u'')t_*} %
    = \contapp{\rbk{t_*}}{\rbk{u_*''}} %
    = \rbk{\contapp{t_*}{u_*''}} %
    = \rbk{u''[t_*/k]}$
  \end{itemize}
\end{proof}

Concluding, we can state that:
\begin{itemize}
\item $<p,t_*>;\lambda <x, k>. u$ is a $\beta$-redex and that
  $<p,t_*>;\lambda <x, k>. u \redto u'[p/x,t_*/k]$ is the
  \emph{$\beta$-rule} of the \ptq-calculus;
\item all the other redexes of the calculus are \emph{control redexes}
  and the corresponding rules are the \emph{control rules} of the
  \ptq-calculus;
\item a reduction $u_*\rredto u_*'$ is a \emph{control reduction} when
  it does not contract any $\beta$-redex.
\end{itemize}

\subsection{Termination of the computations of the \ptq-calculus}

The reduction of any \el-term $u_*$ is deterministic---let us recall
that any \el-term is in normal form or is a redex (the only one in the
term). Therefore, for every \el-term there is only one maximal
reduction sequence that, as we are going to prove, ends with a normal
form.

By Proposition~\ref{prp:rb-soundness}, any reduction of $u_*$ cannot
contain a number of $\beta$-rules greater than the length of the
longest reduction of $\rbk{u_*}$ (let us recall that such a term is
typable in the simply typed $\lambda$-calculus, thus it is strongly
normalizing). Therefore, if the \el-term $u_*$ has an infinite
reduction, such a reduction must eventually end in an infinite
sequence of control reductions.

In order to prove that the control reduction are terminating, we can
associate a measure to every (\tl-closed) \ptq-term that, given a
function from the set of the \pl-variables into the set of the natural
numbers $\NN$, maps
\begin{itemize}
\item every \tl-term into a function of type $(\NN\to\NN)\to\NN\to\NN$;
\item every \pl-term into a natural number;
\item every \ql-term and every \el-term into a function of type
  $(\NN\to\NN)\to\NN$.
\end{itemize}

Denoting by $\missn{a}{\sigma}$ the map that, given a function
$\sigma$ from the set of the \pl-variables into $\NN$, associates to a
\ptq-term $a$ its measure, the maps
$$\missn{t_*}{\sigma}:(\NN\to\NN)\to\NN\to\NN
\qquad
\missn{p}{\sigma}:\NN$$
$$\missn{q}{\sigma}:(\NN\to\NN)\to\NN
\qquad
\missn{u_*}{\sigma}:(\NN\to\NN)\to\NN$$
are defined by
$$\begin{array}{lcl@{\qquad\qquad}lcl}
  \missn{*}{\sigma}\,f\,n & = & f\,n &
  \missn{x}{\sigma} & = & \sigma(x) \\[1.1ex]
  \missn{<p,t_*>}{\sigma}\,f\,n & = & n &
  \missn{\lambda<x,k>.u}{\sigma} & = & 0 \\[1.1ex]
  \missn{\lambda x.u_*}{\sigma}\,f\,n & = & \missn{u_*}{\sigma[x\mapsto n]}(f) &
  \missn{\lambda k.u}{\sigma} & = & \missn{u_*}{\sigma}\,\id \\[1.5ex]
\end{array}
$$
$$\begin{array}{lcl}
  \missn{\clambda k.u}{\sigma}\,f & = & \missn{u_*}{\sigma}\,f
  \\[1.1ex]
  \missn{t_*;p}{\sigma}\,f  &
  = & (\missn{t_*}{\sigma}\,f\,\missn{p}{\sigma}) + 1
  \\[1.1ex]
  \missn{qt_*}{\sigma}\,f &
  = & (\missn{q}{\sigma}\,(\missn{t_*}{\sigma}\,f)) + 1
\end{array}$$

The key properties that have inspired the definition of the above
measure are summarized by the following fact.

\begin{fact}\label{fac:mis-subst}
  Let $a$ be a \tl-closed \tl-term or a \pl-term or a \ql-term or a
  \tl-closed \el-term.
  \begin{enumerate}
  \item $\missn{a[p/x]}{\sigma} =
    \missn{a}{\sigma[x\mapsto\missn{p}{\sigma}]}$, for every \pl-term $p$.
  \item $\missn{\contapp{t_*}{a}}{\sigma}\,f =
    \missn{a}{\sigma}(\missn{t_*}{\sigma}\,f)$, for every \tl-closed
    \tl-term $t_*$, when $a$ is a \tl-closed \tl-term or a
    \tl-closed \el-term.
  \end{enumerate}
\end{fact}
\begin{proof}
  By induction on the structure of $a$.
  \begin{enumerate}
  \item Straightforward.
  \item In the base case $a=*$, we have
    $\missn{\contapp{t_*}{*}}{\sigma}\,f=\missn{t_*}{\sigma}\,f= %
    \missn{*}{\sigma}\,(\missn{t_*}{\sigma}\,f)$. %
    The induction steps are:
    \begin{enumerate}
    \item if $a=<p,t_*'>$, then %
      $\missn{\contapp{t_*}{<p,t_*'>}}{\sigma}\,f = %
      \id = \missn{<p,t_*'>}{\sigma}\,(\missn{t_*}{\sigma}\,f)$; %
    \item if $a=\lambda x.u_*$ with $x\not\in\FV{t_*}$, then %
      $\missn{\contapp{t_*}{\lambda x.u_*}}{\sigma}\,f = %
      \lambda n.\missn{\contapp{t_*}{u_*}}{\sigma[x\mapsto n]}\,f$ $=$ %
      (by the induction hypothesis and $x\not\in\FV{t_*}$) %
      $\lambda n.\missn{u_*}{\sigma[x\mapsto n]}\,
      (\missn{t_*}{\sigma}\,f)$ $= %
      \missn{\lambda x.u_*}{\sigma}\,(\missn{t_*}{\sigma}\,f)$; %
    \item if $a=t_*';p$, then %
      $\missn{\contapp{t_*}{(t_*';p)}}{\sigma}\,f = %
      \missn{\contapp{t_*}{t_*'}}{\sigma}\,f\,\missn{p_*}{\sigma}+1 =$ %
      (by the induction hypothesis) %
      $\missn{t_*'}{\sigma}\,(\missn{t_*}{\sigma}\,f)\,
      \missn{p_*}{\sigma}+1= %
      \missn{t_*';p}{\sigma}\,(\missn{t_*}{\sigma}\,f)$; %
    \item if $a=qt_*'$, then
      $\missn{\contapp{t_*}{(qt_*')}}{\sigma}\,f = %
      \missn{q}{\sigma}\,(\missn{\contapp{t_*}{t_*'}}{\sigma}\,f)+1 =$ %
      (by the induction hypothesis) %
      $\missn{q}{\sigma}(\missn{t_*'}{\sigma}\,
      (\missn{t_*}{\sigma}\,f))+1 = %
      \missn{qt_*'}{\sigma}\,(\missn{t_*}{\sigma}\,f)$. %
    \end{enumerate}
  \end{enumerate}
\end{proof}

The measure of an \el-term $u_*$ corresponds to the length of its
longest control reduction. Let us define
$$\lencntred{u_*}=\sup\{l\mid 
\mbox{$l$ is the length of a control reduction $u_*\rredto u_*'$}\}$$.

\begin{lemma}\label{lem:control-norm}
  For every \el-term, $\lencntred{u_*}$ is finite. Moreover,
  $\lencntred{u_*}=\missn{u_*}{o}\,\id-1$, where $o$ is the map that
  associates $0$ to every free variable of $u_*$.
\end{lemma}
\begin{proof}
  Let us prove by induction on $\lencntred{u_*}$ that
  $\missn{u_*}{o}\,\id=\lencntred{u_*}+1$. We proceed by case
  analysis.
  \begin{enumerate}
  \item Let $u_*=t_*;p$ with $t_*=*$ or $t_*=<p',t_*'>$. We have that
    $\missn{t_*;p}{o}\,\id=\missn{t_*}{o}\,\id\,\missn{p}{o}+1= %
    \missn{p}{o}+1$. %
    Therefore, we have to show that $\missn{p}{o}=\lencntred{t_*;p}$.
    \begin{enumerate}
    \item if $p=x$, then $\missn{x}{o}=o(x)=0=\lencntred{t_*;x}$,
      since $t_*;x$ is a normal form;
    \item if $p=\lambda<k,x>.u_*'$, then
      $\missn{\lambda<k,x>.u_*'}{o}=0=\lencntred{t_*;\lambda<k,x>.u_*'}$
      since $t_*;\lambda<k,x>.u_*'$ is a normal form for the control
      rules;
    \item if $p=\lambda k.u_*'$, then %
      $t_*;\lambda k.u_*'\redto \contapp{t_*}{u_*'}$ %
      and $\missn{\lambda k.u_*'}{o}=\missn{u_*'}{o}\,\id =$ %
      (since $\missn{t_*}{o}\,\id=\id$, by the hypothesis on $t_*$) %
      $\missn{u_*'}{o}\,(\missn{t_*}{o}\,\id) =$ %
      (by Fact~\ref{fac:mis-subst}) %
      $\missn{\contapp{t_*}{u_*'}}{o}\,\id =$ %
      (by the induction hypothesis) %
      $\lencntred{\contapp{t_*}{u_*'}}+1 = %
      \lencntred{t_*;\lambda k.u_*'}$.
    \end{enumerate}
  \item Let $u_*=\lambda x.u_*';p$. We have that %
    $\lambda x.u_*';p\redto u_*'[p/x]$ %
    and $\missn{\lambda x.u_*';p}{o}\,\id= %
    \missn{\lambda x.u_*'}{o}\,\id\,\missn{p}{o}+1 = %
    \missn{u_*'}{o[x\mapsto \missn{p}{o}]}\,\id+1 = $ %
    (by Fact~\ref{fac:mis-subst}) $\missn{u_*'[p/x]}{o}\,\id+1 = $ %
    (by the induction hypothesis) $\lencntred{u_*'[p/x]}+2 = %
    \lencntred{\lambda x.u_*';p}+1$.
  \item Let $u_*=(\clambda k.u_*')t_*$. We have that $(\clambda
    k.u')t_*\redto \contapp{t_*}{u_*'}$ and %
    $\missn{(\clambda k.u')t_*}{o}\,\id = %
    \missn{\clambda k.u'}{o}\,(\missn{t_*}{o}\,\id)+1 = %
    \missn{u_*'}{o}\,(\missn{t_*}{o}\,\id)+1 =$ %
    (by Fact~\ref{fac:mis-subst}) %
    $\missn{\contapp{t_*}{u_*'}}{o}\,\id+1 = $ %
    (by the induction hypothesis) %
    $\lencntred{\contapp{t_*}{u_*'}}+2 = %
    \lencntred{(\clambda k.u_*')t_*}+1$.
  \end{enumerate}
\end{proof}

We can then conclude that the \ptq-calculus is (strongly) normalizing.

\begin{theorem}\label{thm:SN}
  There is no infinite reduction of any \el-term of the \ptq-calculus.
\end{theorem}
\begin{proof}
  By Proposition~\ref{prp:rb-soundness}, the maximal reduction of an
  \el-term $u_*$ cannot contain an infinite number of
  $\beta$-rules. By Lemma~\ref{lem:control-norm}, that reduction
  cannot contain an infinite sequence of control rules
  neither.
\end{proof}

The previous result ensures that the \ptq-calculus may be used as a
computational tool for the implementation of $\beta$-reduction. In
fact, Theorem~\ref{thm:SN} implies that, when the \tl-closed \el-term
$u_*$ is a representation of a given simply typed $\lambda$-term $M$,
the reduction of $u_*$ terminates with a representation of the normal
form of $M$.

\begin{theorem}\label{thm:sim-beta}
  Let $N$ be the normal form of a simply typed $\lambda$-term $M$. If
  $u_*$ is an \el-term s.t.\ $\rbk{u_*}=M$, there is $u_*\mcto u'_*$ s.t.\
  $\rbk{u'_*}=N$. In particular, $\rbk{u'_*}=N$ when $u'_*$ is the
  normal form of $u_*$.
\end{theorem}
\begin{proof}
  By Proposition~\ref{prp:rb-soundness} and Theorem~\ref{thm:SN}.
\end{proof}


\section{Translations}

Theorem~\ref{thm:sim-beta} shows that the \ptq-calculus is a
well-suited target language for the ``compilation'' of
$\lambda$-terms: given a suitable translation of $\lambda$-terms into
\ptq-terms, we can compute the normal form of a $\lambda$-term by
reducing its corresponding \ptq-term; where by suitable translation we
mean a (total) map that inverts the readback.

Let us remind that, since in a \ptq-term there is at most one redex,
the reduction of the \ptq-term is deterministic and induces a
particular reduction strategy of its readback. As a consequence, any
translation of $\lambda$-terms into \ptq-terms defines a reduction
strategy for $\lambda$-terms. In particular, and this is the
interesting computational property of the \ptq-calculus, we can define
translations that implement Call-by-Value and Call-by-Name (see
Plotkin~\cite{Plotk75}).

\subsection{Call-by-Value and Call-by-Name $\lambda$-calculus}

In the Call-by-Value (CbV) and in the Call-by-Name (CbN)
$\lambda$-calculus a $\lambda$-term is a \emph{value} if it is not an
application. The main rule, in both cases, is the $\beta$-rule of
$\lambda$-calculus that, in CbV, is restricted to the case in which a
$\lambda$-abstraction is applied to a value.

In the paper we shall consider the lazy case only, that is we shall
not reduce the $\beta$-redexes that are in the scope of a
$\lambda$-abstraction.

The reduction rules of CbN and CbV will be given by means of inference
rules that do not extend to contexts. In the case of CbN, we have two
rules (small step natural semantics of CbN):
$$
\urule{\phantom{M\nto M_1}}{(\lambda x. M)N\nto M[N/x]}{\beta_n}
\qquad
\urule{M\nto M_1}{MN\nto M_1N}{}
$$ %
The $\beta_n$-rule is the standard $\beta$-reduction of
$\lambda$-calculus restricted to the case in which the reducing term
is a $\beta$-redex. The second rule, instead, allows to reduce a
$\beta$-redex when it is the left-most-outer-most head application of the term.

The CbN-normal form of any closed $\lambda$-term $M$ is a value $V$,
say $M\cbnev V$. The relation $\cbnev$ is defined by the following rules
(big step natural semantics of CbN):
$$\urule{\phantom{M\cbnev M_1}}{V\cbnev V}{}
\qquad
\brule %
   { %
M\cbnev \lambda x. M_1
   } %
   { %
M_1[N/x]\cbnev V
   } %
   { %
MN \cbnev V
   } %
   { } %
$$
where $V$ denotes a value.

The main reduction rule of CbV is the usual $\beta$-rule restricted to
the case in which the reducing term is a $\beta$-redex whose argument
is a value, namely
$$\urule{
}{ (\lambda x. M)V\vto M[V/x] }{\beta_v}$$ 
where $V$ denotes a value.

As in the case of CbN, the small steps natural semantics of CbV is
completed by the inference rules that allow to reduce the head redexes
of an application that, in this case, can be in the argument part
also. However, we have to fix an evaluation order deciding which part
of an application we want to reduce first. The rules that reduce the
function part first are
$$
\urule{M\vto M_1}{MN\vto M_1N}{}
\qquad
\urule{N\vto N_1}{VN\vto VN_1}{}
$$
where $V$ denotes a value. Otherwise, if one wants to reduce the
argument first, the rules are
$$
\urule{N\vto N_1}{MN\vto MN_1}{}
\qquad
\urule{M\vto M_1}{MV\vto M_1V}{}
$$
where $V$ denotes a value.

Both choices lead to the following big step natural semantics for CbV
$$
\urule{\phantom{M\cbnev M_1}}{V\cbnev V}{}
\qquad
\trule %
   { %
M\cbvev \lambda x. M_1
   } %
   { %
M_1[W/x]\cbvev V
   } %
   { %
N\cbvev W
   }%
   { %
MN \cbnev V
   } %
   { } %
$$
where $V$ and $W$ denote values.

The typing rules of CbV and CbN are the usual ones of simply typed
$\lambda$-calculus.

\subsection{Plotkin's translations}
\label{sec:plotkin-trans}

In his seminal paper~\cite{Plotk75}, Plotkin gave two translations
that allow to implement CbN by CbV and vice versa. Both the
translations map an application into a value, that is into a term that
is in normal form---let us remind that we do not reduce in the scope
of an abstraction. In order to start the computation of the translated
term, we have to pass the identity $I=\lambda x.x$ to it---the term
$I$ plays the role of the initial continuation.

\subsubsection{CbN translation}
\label{sec:plotkin-cbn}

The CbN translation is defined by the map
\begin{eqnarray*}
  \plocbn{x} & = & x \\
  \plocbn{\lambda x.M} & = & \lambda k. k (\lambda x.\plocbn{M}) \\
  \plocbn{MN} & = & \lambda k. \plocbn{M} (\lambda m. m \plocbn{N} k) \\
\end{eqnarray*}
that translates a $\lambda$-term $M$ into another $\lambda$-term
$\plocbn{M}$ s.t.\ the CbV reduction of $\plocbn{M}$ corresponds to the
CbN reduction of $MI$.

The untyped CbN translation of terms given above corresponds to the
following translation of typed terms
\begin{eqnarray*}
  \seq{\Gamma}{M:A}
  & \cbnar &
  \seq{\plocbn{\Gamma}}{\plocbn{M}:\plocbn{A}}
\end{eqnarray*}
where, if $\Gamma=x_1:A_1,\ldots,x_n:A_n$, then
$\plocbn{\Gamma}=x_1:\plocbn{A_1},\ldots,x_n:\plocbn{A_n}$, and the
translation of the types $\plocbn{A}$ is defined by
$$
\begin{array}{c@{\qquad\qquad}c}
  \multicolumn{2}{c}{\plocbn{A}\ =\ 
    \auxplocbn{A}\rightarrow o \rightarrow o} \\[+1ex]
  \auxplocbn{X}\ =\ X &
  \auxplocbn{A\to B}\ =\ \plocbn{A}\to\plocbn{B}
\end{array}
$$
where $o$ is a new base type and $X$ denotes a base type.

For the analysis of the correspondence between Plotkin's translation
and the \ptq-translation that we shall give in the following, let us
observe that, by uncurryfying the translation of $A\to B$, we can
assume
\begin{eqnarray*}
  \auxplocbn{A \to B} & = & (\plocbn{A} \times (\auxplocbn{B} \to o)) \to o 
\end{eqnarray*}
Correspondingly, the CbN translation of terms becomes
\begin{eqnarray*}
  \plocbn{x} & = & x \\
  \plocbn{\lambda x.M} & = & \lambda k. k (\lambda (x,h).\plocbn{M} h) \\
  \plocbn{MN} & = & \lambda k. \plocbn{M} (\lambda m. m (\plocbn{N}, k))
\end{eqnarray*}

\subsubsection{CbV translation}
\label{sec:plotkin-cbv}

The CbV translation is defined by the map
\begin{eqnarray*}
  \plocbv{x} & = & \lambda k. k x \\
  \plocbv{\lambda x.M} & = & \lambda k. k (\lambda x. \plocbv{M}) \\
  \plocbv{MN} & = &
  \lambda k. \plocbv{M} (\lambda m. \plocbv{N} (\lambda n. m n k)) \\
\end{eqnarray*}
that translates every $\lambda$-term $M$ into another $\lambda$-term
$\plocbv{M}$ s.t.\ the CbN reduction of $\plocbv{M}$ corresponds to the
CbV reduction of $MI$.

The map above ensures that, in an application $MN$, the function $M$
is evaluated first. Replacing the translation of $MN$ with %
$$\plocbv{MN} = %
  \lambda k. \plocbv{N} (\lambda n. \plocbv{M} (\lambda m. m n k)) %
$$ %
we get the translation for the case in which, in an application $MN$,
the argument $N$ is evaluated first.

The untyped CbV translation of terms given above corresponds to the
following translation of typed terms
\begin{eqnarray*}
  \seq{\Gamma}{M:A}
  & \cbvar &
  \seq{\auxplocbv{\Gamma}}{\plocbv{M}:\plocbv{A}}
\end{eqnarray*}
where, if $\Gamma=x_1:A_1,\ldots,x_n:A_n$, then
$\auxplocbv{\Gamma}=x_1:\auxplocbv{A_1},\ldots,x_n:\auxplocbv{A_n}$ and the
translations of types $\auxplocbv{A}$ and $\plocbv{A}$ are defined by
$$
\begin{array}{c@{\qquad\qquad}c}
  \multicolumn{2}{c}{\plocbv{A}\ =\ 
    \auxplocbv{A}\rightarrow o \rightarrow o} \\[+1ex]
  \auxplocbn{X}\ =\ X & 
  \auxplocbn{A\to B}\ =\ \auxplocbv{A}\to\plocbv{B}
\end{array}
$$
where $o$ is a new base type and $X$ denotes a base type.

As in the case of CbN translation, by uncurryfying the translation of
$A\to B$, we get
\begin{eqnarray*}
  \auxplocbn{A \to B} & = & (\auxplocbv{A} \times (\auxplocbv{B} \to o)) \to o 
\end{eqnarray*}
Correspondingly, the CbV translation of terms becomes
\begin{eqnarray*}
  \plocbv{x} & = & \lambda k. k x \\
  \plocbv{\lambda x.M} & = & \lambda k. k (\lambda (x,h). \plocbv{M} h) \\
  \plocbv{MN} & = &
  \left\{
  \begin{array}{l}
    \lambda k. \plocbv{M} (\lambda m. \plocbv{N} (\lambda n. m (n,k)))
    \qquad\quad \mbox{eval $M$ first}\\
    \lambda k. \plocbv{N} (\lambda n. \plocbv{M} (\lambda m. m (n,k)))    
    \qquad\quad \mbox{eval $N$ first}
  \end{array}
  \right .
  \\
\end{eqnarray*}
where, for $MN$, we have to choose the upper translation if we want to
evaluate $M$ first or the lower translation if we want to
evaluate $N$ first.

\subsection{Call-by-Name \ptq-translation}

The CbN \ptq-translation derives from the translation that maps every
sequent derivable in minimal logic into the corresponding
\ptq-sequent, by translating every formula of minimal logic into a
\pl-formula, that is
$$\Gamma \vdash A\qquad \cbnar \qquad
  \pf{\Gamma}\vdash\pf{A}
$$%
The simplest way to get the above correspondence is by the
CbN-translation in Figure~\ref{fig:ptq-trans}.

\begin{figure}[tbp]
  \centering
  \begin{minipage}[t]{.45\linewidth}
    \centering
    Call-by-Name
    \vspace{-1ex}
    \begin{eqnarray*}
      \cbn{x} & = & x \\
      \cbn{\lambda x. M} & = & \lambda<x,k>.k;\cbn{M} \\
      \cbn{M N} & = & \lambda k. <\cbn{N},k>;\cbn{M}
    \end{eqnarray*}
  \end{minipage}
  \begin{minipage}[t]{.45\linewidth}
    \centering
    Call-by-Value
    \vspace{-1ex}
    \begin{eqnarray*}
      \cbv{x} & = & \clambda k. k;x \\
      \cbv{\lambda x. M} & = & \clambda k. k;(\lambda<x,k>.\cbv{M}k)\\
      \cbv{M N} & = & \clambda k. \cbv{N}(\lambda x.\cbv{M}<x,k>)
    \end{eqnarray*}
  \end{minipage}
  \caption{\ptq-translations}
  \label{fig:ptq-trans}
\end{figure}

\begin{proposition}
  Let $\Gamma \vdash M:A$ be derivable in the simply typed
  $\lambda$-calculus. Then, $\pf{\Gamma}\vdash\cbn{M}:\pf{A}$ is
  derivable in the \ptq-calculus.
\end{proposition}
\begin{proof}
  By induction on the structure of $M$.
\end{proof}

\subsubsection{Correspondence with Plotkin's CbN translation}

In order to relate the CbN \ptq-translation with Plotkin's CBN
translation, let us observe that we may map the translated Plotkin's
types according to the following schema:
$$\plocbn{A} \quad = \quad \auxplocbn{A} \to o \to o %
\quad \trar \quad \pf{A}$$ %
$$\auxplocbn{A}\quad\trar\quad\pf{A}\qquad\qquad\qquad %
\auxplocbn{A}\to o\quad\trar\quad\tf{A}%
$$ %
that also implies, in the translation with pairs,
$$\auxplocbn{A\to B}\to o \quad = \quad %
\plocbn{A}\times (\auxplocbn{B}\to o) 
\quad \trar \quad \tf{A \to B} %
$$ %
Correspondingly, the CbN Plotkin's translation of terms becomes
\begin{eqnarray*}
  \plocbn{x} & \trar & x \\
  \plocbn{\lambda x.M} & \trar & \lambda k. k ; (\lambda (x,m).m ;\cbn{M}) \\
  \plocbn{MN} & \trar & \lambda k. (\lambda m. <\cbn{N}, k> ; m) ; \cbn{M}
\end{eqnarray*}
Such a translation can be simplified to the \ptq-translation in
Figure~\ref{fig:ptq-trans} by observing that, by $\eta$-equivalence
$\lambda k. (\lambda m. <\plocbn{N}, k> ; m) ; \plocbn{M} = \lambda
k. <\plocbn{N}, k> ; \plocbn{M}$ and that, since in the translation we
never use \tl-terms with the shape $\lambda x.u$, we can also assume
that $ \lambda k. k ; (\lambda <x,m>.m ;\plocbn{M})$ is equivalent to
$\lambda <x,m>.m ;\plocbn{M}$.

\subsection{Call-by-Value \ptq-translation}

The CbV \ptq-translation derives from the translation of minimal logic
that maps the premises of the ending sequent into \pl-formulas and its
conclusion into a \ql-formula
$$\Gamma \vdash A\qquad \trar \qquad
    \pf{\Gamma}\vdash\qf{A}$$
In order to get the above correspondence, one can easily find the
CbV translation in Figure~\ref{fig:ptq-trans}.

\begin{proposition}
  Let $\Gamma \vdash M:A$ be derivable in the simply typed
  $\lambda$-calculus. Then, $\pf{\Gamma}\vdash\cbv{M}:\qf{A}$ is
  derivable in the \ptq-calculus.
\end{proposition}
\begin{proof}
  By induction on the structure of $M$.
\end{proof}

\subsubsection{Correspondence with Plotkin's CbV translation}

In order to relate the CbV \ptq-translation with Plotkin's CbV
translation, the translated Plotkin's types can be mapped to
\ptq-types according to the following schema:
$$\plocbn{A} \quad = \quad \auxplocbn{A} \to o \to o %
\quad \trar \quad \qf{A}$$ %
$$\auxplocbn{A}\quad\trar\quad\pf{A}\qquad\qquad\qquad %
\auxplocbn{A}\to o\quad\trar\quad\tf{A}%
$$ %
that also implies, in the translation with pairs,
$$\auxplocbn{A\to B}\to o \quad = \quad %
\auxplocbn{A}\times (\auxplocbn{B}\to o) 
\quad \trar \quad \tf{A \to B} %
$$ %
Correspondingly, the CbV Plotkin's translation of terms becomes
\begin{eqnarray*}
  \plocbv{x} & \trar & \clambda k. k ; x \\
  \plocbv{\lambda x.M} & \trar & \clambda k. k ; (\lambda <x,h>. h ; \cbv{M}) \\
  \plocbv{MN} & \trar &
  \left\{
  \begin{array}{l}
    \clambda k. \cbv{M} (\lambda m. (\cbv{N} (\lambda n. <n,k> ; m))) \\
    \clambda k. \cbv{N} (\lambda n. (\cbv{M} (\lambda m. <n,k> ; m)))
  \end{array}
  \right .
\end{eqnarray*}
The translation in which $M$ is evaluated first cannot be
simplified. The case in which $N$ is evaluated first, instead, can be
reduced to the CbV \ptq-translation in Figure~\ref{fig:ptq-trans}
because of the $\eta$-equivalence $\lambda m. <n,k> ; m = <n,k>$.

\section{Properties of the \ptq-translations}

\subsection{Precomputation}
\label{sec:precomp}

The \ptq-translations map $\lambda$-terms into \pl-terms that, in the
\ptq-calculus, are irreducible. In order to eval a translated term we
have to combine it with a test. The natural choice for such an initial
test is the constant $*$ that, as already remarked, in our framework
plays the role of the initial continuation.

By reducing the \ptq-translation $u_*$ of a $\lambda$-term $M$, we do
not get the translation $u_*'$ of some reduct $N$ of $M$. We shall see
instead that we can get some $u_*''$ s.t.\ $\rbk{u_*''}=N$, which
differs from $u_*'$ for the reduction of some control redexes, namely
$u_*' \rredto u_*''$ by a sequence of control rules. Since we know
that the readback $M=\rbk{u_*}$ of an \el-term $u_*$ is not changed by
the control rules (Proposition~\ref{prp:rb-soundness}) and that the
control reductions are terminating (Lemma~\ref{lem:control-norm}), we
can take the normal form of $u_*$ for the control rules as another
standard representation of $M$. The computation of such a normal form
can be merged with the \ptq-translations by defining two variants of
the \ptq-translations that map $\lambda$-terms into \el-terms.

The Call-by-Name translation from $\lambda$-terms to \el-terms is
defined by:
\begin{center}
  \vspace{-2em}
  \begin{minipage}[t]{.45\linewidth}
    \begin{eqnarray*}
      \varcbn{V} & = & *;\auxcbn{V} \\
      \varcbn{M N} & = & \contapp{<\cbn{N},*>}{\varcbn{M}}
    \end{eqnarray*}
  \end{minipage}
  \begin{minipage}[t]{.45\linewidth}
    \begin{eqnarray*}
      \auxcbn{x} & = & x \\
      \auxcbn{\lambda x.M} & = & \lambda<x,k>.k;\cbn{M}
    \end{eqnarray*}
  \end{minipage}
\end{center}
where $V$ is a value. The Call-by-Value translation from
$\lambda$-terms to \el-terms is defined by:
\begin{center}
  \vspace{-2em}
  \begin{minipage}[t]{.45\linewidth}
    \begin{eqnarray*}
      \varcbv{V} & = & *;\auxcbv{V} \\
      \varcbv{M V} & = & %
      \contapp{<\auxcbv{V},*>}{\varcbv{M}} \\
      \varcbv{M N} & = & %
      \contapp{(\lambda x.\cbv{M}<x,*>)}{\varcbv{N}} %
    \end{eqnarray*}
  \end{minipage}
  \begin{minipage}[t]{.45\linewidth}
    \begin{eqnarray*}
      \auxcbv{x} & = & x \\
      \auxcbv{\lambda x.M} & = & \lambda<x,k>.\cbv{M}k
    \end{eqnarray*}
  \end{minipage}
\end{center}
where $V$ is a value and $N$ is not a value.


For every value $V$,
$$\cbn{V}=\lambda k.k;\auxcbn{V}
\qquad\qquad\qquad
\cbv{V}=\clambda k.k;\auxcbv{V}$$

\begin{remark}
  \label{rem:var-ptq-tr}
  Let us define
  \begin{eqnarray*}
    [p_1,\ldots,p_k] & = &
    \begin{cases}
      <\cbn{p_1},<\ldots,<\cbn{p_k},*>\ldots>>
      & \mbox{for $k\gt0$} \\
      * & \mbox{for $k=0$}
    \end{cases}
  \end{eqnarray*}
  \begin{enumerate}
  \item\label{rem:var-ptq-tr:item:n} $\varcbn{M\,M_1 \ldots M_k} = %
    \contapp{[\cbn{M_1},\ldots,\cbn{M_k}]}{\auxcbn{M}}$.
    In particular, when $M=V$ is a value
    \begin{eqnarray*}
      \varcbn{V\,M_1 \ldots M_k} & = & %
      [\cbn{M_1},\ldots,\cbn{M_k}];\auxcbn{V} %
    \end{eqnarray*}
  \item\label{rem:var-ptq-tr:item:v} $\varcbv{M\,V_1\ldots V_k} = %
    \contapp{[\cbv{V_1},\ldots,\cbv{V_k}]}{\auxcbv{M}}$%
    where $V_1,\ldots,V_k$ are values.  In particular, when $M=V$ is a
    value and when $M=PQ$ where $Q$ is not a value
    \begin{eqnarray*}
      \varcbv{V\,V_1\ldots V_k} & = & %
      [\auxcbv{V_1},\ldots,\auxcbv{V_k}];\auxcbv{V} \\ %
      \varcbv{P Q V_1\ldots V_k} & = & %
      \contapp{(\lambda x.\cbv{P} %
        [x,\auxcbv{V_1},\ldots,\auxcbv{V_k}])} %
      {\varcbv{Q}} %
    \end{eqnarray*}
  \end{enumerate}
\end{remark}


\begin{lemma}
  \label{lem:pre-comp}
  For every $\lambda$-term $M$,
  \begin{enumerate}
  \item\label{lem:pre-comp:item:1} both $\varcbn{M}$ and $\varcbv{M}$
    are in normal form for the control rules;
  \item\label{lem:pre-comp:item:2} by a sequence of control rules  
    \begin{enumerate}
    \item $*;\cbn{M}\rredto\varcbn{M}$,
    \item $\cbv{M}*\rredto\varcbv{M}$.
    \end{enumerate}
  \end{enumerate}
\end{lemma}
\begin{proof}
  Let us separately prove the two items of the statement.
  \begin{enumerate}
  \item Let us start by proving the following claim.

    \begin{uclaim}
      There are two values $V^n$ and $V^v$ and two \tl-closed terms
      $t_*^n$ and $t_*^v$ s.t.\ $\varcbn{M}=t_*^n;\auxcbn{V^n}$ and
      $\varcbv{M}=t_*^v;\auxcbv{V^v}$, and s.t.:
      \begin{enumerate}
      \item when $M$ is a value, $t_*^n = t_*^v=*$ and $M=V^n=V^v$;
      \item when $M$ is not a value, $t_*^n$ and $t_*^v$ are pairs,
        namely $t_*^n=<p',t_*'>$ for some $p'$ and $t_*'$ and
        $t_*^v=<p'',t_*''>$ for some $p''$ and $t_*''$.
      \end{enumerate}
    \end{uclaim}  
    \emph{Proof of the claim.}  For $\varcbn{M}$, the proof
    immediately follows by item~\ref{rem:var-ptq-tr:item:n} of
    Remark~\ref{rem:var-ptq-tr}: take $M=V^n\,M_1\ldots M_k$. For
    $\varcbv{M}$, the proof exploits the inductive definition of
    $\varcbv{M}$ in item~\ref{rem:var-ptq-tr:item:v} of
    Remark~\ref{rem:var-ptq-tr}: the base case is immediate, just take
    $M=V^v\,V_1\ldots V_K$; the induction step $M=PQV_1\ldots V_k$
    holds by the induction hypothesis on $\varcbv{Q}$ and by the fact
    that $Q$ is not a value. $\square$

    Then, in order to conclude that $\varcbn{M}=t_*^n;\auxcbn{V^n}$
    and $\varcbv{M}=t_*^v;\auxcbv{V^v}$ are in normal form for the
    control rules, let us observe that, for any value $V$:
    \begin{enumerate}
    \item if $V=x$, then $\auxcbn{V}=\auxcbv{V}=x$;
    \item if $V=\lambda x.N$, then $\auxcbn{V}=\lambda<x,k>.u^n$ and
      $\auxcbv{V}=\lambda<x,k>.u^v$ for some $u^n$ and $u^v$.
    \end{enumerate}
  \item By structural induction on $M$. When $M=V$, where $V$ is a
    value (base case), it is readily seen that
    $*;\cbn{V}\redto*;\auxcbn{V}$ and $\cbv{V}*\redto*;\auxcbv{V}$ by
    a control reduction. When $M=PQ$, by the induction hypothesis, we
    have three control reductions s.t.: $*;P\rredto\varcbn{P}$, for
    the CbN; $\cbv{P}*\rredto\varcbv{P}$ and $\cbv{Q}*\rredto
    \varcbv{Q}$, for the CbV. Therefore, by Lemma~\ref{lem:red-subst},
    \begin{itemize}
    \item for the CbN, we have the control reduction \\ %
      $*;\cbn{M}=*;\cbn{PQ}\redto<\cbn{Q},*>;\cbn{P} %
      \rredto\contapp{<\cbn{Q},*>}{\varcbn{P}}=\varcbn{M}$;
    \item for the CbV,
      \begin{enumerate}
      \item when $Q=V$ is a value, we have the control reduction \\ %
        $*;\cbv{M}=\cbv{PV}*=\contapp{<\auxcbv{V},*>}{\cbv{P}}\rredto %
        \contapp{<\auxcbv{V},*>}{\varcbv{P}}=\varcbv{M}$; %
      \item when $Q$ is not a value, we have the control reduction \\ %
        $*;\cbv{M}=\cbv{PQ}*\redto %
        \contapp{(\lambda x.\cbv{P}<x,*>)}{\cbv{Q}}\rredto %
        \contapp{(\lambda x.\cbv{P}<x,*>)}{\varcbv{Q}}=\varcbv{M}$. %
      \end{enumerate}
    \end{itemize}
  \end{enumerate}
\end{proof}

\subsection{Readback}
\label{sec:prop-rbk}

One of the key properties of the \ptq-translations is that the
readback of a translated term is the term itself.

\begin{proposition}
  For every $\lambda$-term $M$,
  \begin{enumerate}
  \item $\rbk{*;\cbn{M}} = \rbk{\varcbn{M}} = \rbk{\cbn{M}} = M$
  \item $\rbk{\cbv{M}*} = \rbk{\varcbv{M}} = \rbk{\cbv{M}} = M$
  \end{enumerate}
\end{proposition}
\begin{proof}
  By the definition of readback, it is readily seen that
  $\rbk{*;\cbn{M}} = \rbk{\cbn{M}}$ and that $\rbk{\cbv{M}*} =
  \rbk{\cbv{M}}$.
  \begin{enumerate}
  \item We shall prove $\rbk{\cbn{M}} = M$, by induction on the
    structure of $M$.
    \begin{enumerate}
    \item $\rbk{\cbn{x}} = x$
    \item $\rbk{\cbn{\lambda x.P}} =$ %
      $\rbk{\lambda <x,k>. k; \cbn{P}} =
      \lambda x.\rbk{*;\cbn{P}} = $ %
      (by the induction hypothesis) %
      $\lambda x. P$ %
    \item $\rbk{\cbn{PQ}} = $ %
      $\rbk{\lambda k.<\cbn{Q},k>;\cbn{P}} =
      \contapp{\rbk{<\cbn{Q},*>}}{\rbk{\cbn{P}}} =
      \rbk{\cbn{P}}\rbk{\cbn{Q}} = $ %
      (by the induction hypothesis) $PQ$ %
    \end{enumerate}
  \item By induction on the structure of $M$, we shall prove that
    $\rbk{\cbv{M} t_*} = \contapp{\rbk{t_*}}{M}$.
    \begin{enumerate}
    \item $\rbk{\cbv{x}t_*} = \contapp{\rbk{t_*}}\rbk{\clambda k.k;x} =
      \contapp{\rbk{t_*}}\rbk{*;x} = \contapp{\rbk{t_*}}{x}$
    \item $\rbk{\cbv{PQ}t_*} = %
      \contapp{\rbk{t_*}}{\rbk{\cbv{Q}(\lambda
          x.\cbv{P}<x,*>)}} = $ %
      (by the induction hypothesis) %
      $\contapp{\rbk{t_*}}{\contapp{\rbk{\lambda
            x.\cbv{P}<x,*>}}{Q}}= 
      \contapp{\rbk{t_*}}{\rbk{\cbv{P}<x,*>}[Q/x]}= $ %
      (by the induction hypothesis) %
      $\contapp{\rbk{t_*}}{(\contapp{\rbk{<x,*>}}{P})[Q/x]} =
      \contapp{\rbk{t_*}}{Px[Q/x]} = \contapp{\rbk{t_*}}{PQ}$%
    \item $\rbk{\cbv{\lambda x.P}t_*} = %
      \contapp{\rbk{t_*}}{\rbk{\lambda<x,k>.\cbv{P}k}} = %
      \contapp{\rbk{t_*}}{\lambda x.\rbk{\cbv{P}*}} = $ %
      (by the induction hypothesis) %
      $\contapp{\rbk{t_*}}{\lambda x.P}$
    \end{enumerate}
    In particular, $\rbk{\cbv{M}*}=\contapp{\rbk{*}}{M}=M$.
  \end{enumerate}
  By Lemma~\ref{lem:pre-comp}, $*;\cbn{M}\rredto\varcbn{M}$ and
  $\cbv{M}*\rredto\varcbv{M}$ by control reductions. Therefore, by
  Proposition~\ref{prp:rb-soundness}, $\rbk{\varcbn{M}}=\rbk{*;\cbn{M}}=M$
  and $\rbk{\varcbv{M}}=\rbk{\cbv{M}*}=M$.
\end{proof}

\subsection{Soundness and Completeness}
\label{sec:sound-comp}

\begin{lemma}
  \label{lem:tr-subst}
  For every pair of $\lambda$-terms $M,N$ and every value $V$
 \begin{enumerate}
  \item $\cbn{M[N/x]} = \cbn{M}[\cbn{N}/x]$
  \item $\cbn{M[V/x]} = \cbn{M}[\auxcbv{V}/x]$
  \end{enumerate}
\end{lemma}
\begin{proof}
  By induction on $M$.
\end{proof}

\begin{proposition}
\label{prp:simulation}
  For every $\lambda$-term $M$.
  \begin{enumerate}
  \item $M\redto N$ in the CbN $\lambda$-calculus iff
    $\varcbn{M} \rredto \varcbn{N}$.
  \item $M\redto N$ in the CbV $\lambda$-calculus iff
    $\varcbv{M} \rredto \varcbv{N}$.
  \end{enumerate}
\end{proposition}
\begin{proof}
  By induction on $M$. If $M=V$ is a value (the base of the
  induction), it is readily seen that $\varcbn{V}=*;\auxcbn{V}$ and
  $\varcbv{V}=*;\auxcbv{V}$ are in normal form (see the proof of
  item~\ref{lem:pre-comp:item:2} of Lemma~\ref{lem:pre-comp}). Then,
  let us prove the inductive steps of the two items in the statement.
  \begin{enumerate}
  \item Let $M = V\,M_1\ldots M_k$, where $V$ is a value and $k\gt 0$.
    By Remark~\ref{rem:var-ptq-tr}, we know that
    $\varcbn{M}=[\cbn{M_1},\ldots,\cbn{M_k}];\auxcbv{V}$. If $V=x$ is
    a variable, $M$ is a CbN normal form and
    $[\cbn{M_1},\ldots,\cbn{M_k}];x$ is a normal form too. %
    If $V=\lambda x.P$ is a $\lambda$-abstraction, then $M\redto
    P[M_1/x] M_2\ldots M_k=N$ in the CbN and
    \begin{align*}
      \varcbn{M} &=\;
      [\cbn{M_1},\ldots,\cbn{M_k}];\lambda<x,k>.k;\cbn{P} \\
      & \redto\;
      [\cbn{M_2},\ldots,\cbn{M_k}];\cbn{P}[\cbn{M_1}/x] \\
      & =\; [\cbn{M_2},\ldots,\cbn{M_k}];\cbn{P[M_1/x]}
      & \mbox{(by Lemma~\ref{lem:tr-subst})} \\
      & \rredto\;
      \contapp{[\cbn{M_2},\ldots,\cbn{M_k}]}{\varcbn{P[M_1/x]}}
      & \mbox{(by Lemma~\ref{lem:pre-comp} and Lemma~\ref{lem:red-subst})} \\
      & =\; \varcbn{P[M_1/x]M_2\ldots M_k} =\;\varcbn{N}
    \end{align*}
  \item We have to analyze two cases.
    \begin{enumerate}
    \item Let $M = V\,V_1\ldots V_k$, where $V,V_1,\ldots,V_k$ are
      values and $k\gt 0$. By Remark~\ref{rem:var-ptq-tr} we know that
      $\varcbn{M}=[\auxcbv{V_1},\ldots,\auxcbv{V_k}];\auxcbv{V}$. If
      $V=x$ is a variable, $M$ is a CbN normal form and
      $[\auxcbv{V_1},\ldots,\auxcbv{V_k}];x$ is a normal form too. %
      If $V=\lambda x.P$ is a $\lambda$-abstraction, in the CbV %
      $M\redto P[V_1/x] V_2\ldots V_k$ and
      \begin{align*}
        \varcbv{M} &=
        [\auxcbv{V_1},\ldots,\auxcbv{V_k}];\lambda<x,k>.k;\cbv{P} \\
        & \redto
        [\auxcbv{V_2},\ldots,\auxcbv{V_k}];\cbv{P}[\auxcbv{V_1}/x] \\
        & = [\auxcbv{V_2},\ldots,\auxcbv{V_k}];\cbv{P[V_1/x]}
        & \mbox{(by  Lemma~\ref{lem:tr-subst})} \\
        & \rredto 
        \contapp{[\auxcbv{V_2},\ldots,\auxcbv{V_k}]}{\varcbv{P[V_1/x]}}
        & \mbox{(by Lemma~\ref{lem:pre-comp} and Lemma~\ref{lem:red-subst})} \\
        & = \varcbv{P[V_1/x]V_2\ldots V_k}=\;\varcbv{N}
      \end{align*}
    \item Let $M=PQV_1\ldots V_k$, where $V_1,\ldots,V_k$ are values
      and $Q$ is not a value. By Remark~\ref{rem:var-ptq-tr}, %
      $\varcbv{M} = %
      \contapp{(\lambda x.\cbv{P}
        [x,\auxcbv{V_1},\ldots,\auxcbv{V_k}])} %
      {\varcbv{Q}}$. %
      If $Q$ is in normal form for the CbV, then $M$ is in normal form
      for the CbV. By the induction hypothesis, $\varcbv{Q}$ is in
      normal form and (see the proof of Lemma~\ref{lem:pre-comp}) has
      not the shape $*;p$ for some $p$; therefore, $\varcbv{M}$ is in
      normal form. If $Q\redto Q'$ in the CbV, then $M\redto
      PQ'V_1\ldots V_k=N$ in the CbV. By the induction hypothesis,
      $\varcbv{Q}\redto\varcbv{Q'}$ and $\varcbv{M} \redto
      \contapp{(\lambda x.\cbv{P}
        [x,\auxcbv{V_1},\ldots,\auxcbv{V_k}])} {\varcbv{Q'}}$, by
      Lemma~\ref{lem:red-subst}. Then, if $Q'$ is not a value,
      $\contapp{(\lambda x.\cbv{P}
        [x,\auxcbv{V_1},\ldots,\auxcbv{V_k}])}
      {\varcbv{Q'}}=\varcbv{N}$, otherwise, if $Q'=V$ is a value, we
      have
      \begin{align*}
        \varcbv{M} &\redto\; (\lambda
        x.\cbv{P} [x,\auxcbv{V_1},\ldots,\auxcbv{V_k}]);\auxcbv{V} \\
        &\redto\; \cbv{P} [\auxcbv{V},\auxcbv{V_1},\ldots,\auxcbv{V_k}] \\
        &\rredto\;
        [\auxcbv{V},\auxcbv{V_1},\ldots,\auxcbv{V_k}];\varcbv{P} \\ 
        &=\; \varcbv{PVV_1\ldots V_k}=\;\varcbv{N}
      \end{align*}
    \end{enumerate}
  \end{enumerate}
\end{proof}

\begin{theorem}
\label{thm:completeness}
  For every $\lambda$-term $M$.
  \begin{enumerate}
  \item If $M\rredto N$ in the CbN $\lambda$-calculus, then
    $\cbn{M} \rredto \varcbn{N}$, namely
    \begin{displaymath}
      \xymatrix{%
        *++{*;\cbn{M}}
        \ar@{.>}[r]^-{*} &
        *++{\varcbn{N}}  \\
        *++{M} \ar@{->}[r]^-{*}_-{CbN}
        \ar@{|->}|-{\cbn{(\cdot)}}[u]
        & *++{N}\ar@{|->}|-{\varcbn{(\cdot)}}[u]
      }
    \end{displaymath}
  \item If $M \rredto N$ in the CbV $\lambda$-calculus, then
    $\cbv{M} \rredto \varcbv{N}$, namely
    \begin{displaymath}
      \xymatrix{%
        *++{\cbv{M}*}
        \ar@{.>}[r]^-{*} &
        *++{\varcbv{N}}  \\
        *++{M} \ar@{->}[r]^-{*}_-{CbV}
        \ar@{|->}|-{\cbv{(\cdot)}}[u]
        & *++{N}\ar@{|->}|-{\varcbv{(\cdot)}}[u]
      }
    \end{displaymath}
  \end{enumerate}
\end{theorem}
\begin{proof}
  By Lemma~\ref{lem:pre-comp} and Proposition~\ref{prp:simulation}.
\end{proof}

\begin{theorem}
  \label{thm:soundness}
  For every $\lambda$-term $M$.
  \begin{enumerate}
  \item If $*;\cbn{M} \rredto u_*$, then $M \rredto \rbk{u_*}$ in the CbN
    $\lambda$-calculus, namely
    \begin{displaymath}
      \xymatrix{%
        *++{*;\cbn{M}} \ar@{|->}[d]|-{\rbk{\cdot}}
        \ar@{->}[r]^{*} & 
        *++{u_*} \ar@{|->}[d]|-{\rbk{\cdot}} \\
        *++{M} \ar@{->}[r]^{*}_{CbN}
        & *++{\rbk{u_*}}
      }
    \end{displaymath}
  \item If $\cbv{M}* \rredto u_*$, then $M \rredto \rbk{u_*}$ in the CbV
    $\lambda$-calculus, namely

    \begin{displaymath}
      \xymatrix{%
        *++{\cbv{M}*} \ar@{|->}[d]|-{\rbk{\cdot}}
        \ar@{->}[r]^{*} & 
        *++{u_*} \ar@{|->}[d]|-{\rbk{\cdot}} \\
        *++{M} \ar@{->}[r]^{*}_{CbV}
        & *++{\rbk{u_*}}
      }
    \end{displaymath}
  \end{enumerate}
\end{theorem}
\begin{proof}
  By Lemma~\ref{lem:pre-comp}, $\ptqtr{M}\rredto\varptqtr{M}$ by a
  control reduction for CbN and CbV. Therefore, $\rbk{u_*}=M$, for
  every $\ptqtr{M}\rredto u_*\rredto\varptqtr{M}$ (by
  Proposition~\ref{prp:rb-soundness}). By Lemma~\ref{lem:pre-comp},
  $\varptqtr{M}$ is either a normal form or a $\beta$-redex. When
  $\varptqtr{M}$ is a $\beta$-redex, there is a reduction
  $\varptqtr{M}\redto u_*' \rredto u_* \rredto \varptqtr{N}$ s.t.\ all
  the rules but the first one are control rules (see the proof of
  Proposition~\ref{prp:simulation}). By
  Proposition~\ref{prp:rb-soundness}, we have then $M \redto \rbk{u_*'}
  = \rbk{u_*} = N$. Proposition~\ref{prp:simulation} ensures that
  $M\redto\rbk{u_*}$ by CbN or CbV according to the case that we are
  considering.
\end{proof}


\section{Conclusions and further work}

Starting from the notion of test introduced by Girard in
\cite{Gir:Meaning1:99}, we have proposed a new calculus, the
\ptq-calculus, in which we reformulate in logical terms the well-known
duality programs/continuations, namely in terms of the proofs/tests
duality. In the core of the paper we have shown that the \ptq-calculus
has interesting logical and computational properties and, by encoding
$\lambda$-calculus Call-by-Value and Call-by-Name into it, we have
shown that it might be a fruitful framework for the analysis of
reduction strategies and of sequential features of functional
programming languages.

In spite of the classical flavour of \ptq-calculus, in the paper we
have restricted our analysis to the intuitionistic case---mainly
beacuse our goal was to present the \ptq-calculus as a tool for the
study of $\lambda$-calculus Call-by-Value and Call-by-Name. The
natural extension of the analysis pursued in the paper to classical
logic leads to relate our approach to Parigot's $\lambda\mu$-calculus
\cite{Par92,Par97}. In particular, there is a natural bijection
between \ptq-calculus and $\lambda\mu$-calculus that, however, does
not give a simulation, namely the reductions of the \ptq-calculus are
not sound w.r.t.\ the reductions of the $\lambda\mu$-calculus proposed
by Parigot. Such a mismatch reflects the fact that the \ptq-calculus
is neither Call-by-Value nor Call-by-Name, while with the reduction
rules of Parigot the $\lambda\mu$-calculus is essentially
Call-by-Name. Therefore, in order to extend our analysis to the
classical case, we aim at relating the \ptq-calculus with both the
original Call-by-Name $\lambda\mu$-calculus proposed by Parigot and to
the Call-by-Value $\lambda\mu$-calculus proposed by Ong and
Stewart~\cite{ong97curryhoward}, and with Curien and Herbelin's
$\bar{\lambda}\mu\tilde{\mu}$-calculus~\cite{CurHerb00}.

\bibliography{references}

\begin{thebibliography}{10}

\bibitem{CurHerb00}
Pierre-Louis Curien and Hugo Herbelin.
\newblock The duality of computation.
\newblock In {\em ICFP '00: Proceedings of the fifth ACM SIGPLAN international
  conference on Functional programming}, pages 233--243, New York, NY, USA,
  2000. ACM Press.

\bibitem{Dan:EvalContContRestComp:04}
Olivier Danvy.
\newblock On evaluation contexts, continuations, and the rest of the
  computation.
\newblock In Hayo Thielecke, editor, {\em Proceedings of the Fourth {ACM}
  {SIGPLAN} Continuations Workshop {(CW'04)}}, number CSR-040-1 in Technical
  Report. Proceedings of the Fourth {ACM} {SIGPLAN} Continuations Workshop
  {(CW'04)}, Birmingham, UK, 2004.

\bibitem{degr98}
Philippe de~Groote.
\newblock An environment machine for the lambda-mu-calculus.
\newblock {\em Mathematical Structures in Computer Science}, 8(6):637--669,
  1998.

\bibitem{FellFried86}
Matthias Felleisen, Daniel~P. Friedman, Eugene~E. Kohlbecker, and Bruce~F.
  Duba.
\newblock Reasoning with continuations.
\newblock In Albert Meyer, editor, {\em Proceedings of the First Annual IEEE
  Symp. on Logic in Computer Science, {LICS} 1986}, pages 131--141. IEEE
  Computer Society Press, June 1986.

\bibitem{Fisch72}
Michael~J. Fischer.
\newblock Lambda calculus schemata.
\newblock In {\em Proceedings of ACM conference on Proving assertions about
  programs}, pages 104--109, New York, NY, USA, 1972. ACM Press.

\bibitem{FuhThi04}
Carsten F{\"u}hrmann and Hayo Thielecke.
\newblock On the call-by-value {CPS} transform and its semantics.
\newblock {\em Inform. and Comput.}, 188(2):241--283, 2004.

\bibitem{Gir:Meaning1:99}
Jean-Yves Girard.
\newblock On the meaning of logical rules i: syntax vs. semantics.
\newblock In U.~Berger and H.~Schwichtenberg, editors, {\em Computational
  Logic}, volume 165 of {\em NATO series F}, pages 215--272. Springer, 1999.

\bibitem{Griff90}
Timothy~G. Griffin.
\newblock The formulae-as-types notion of control.
\newblock In {\em Conf.\ Record 17th Annual {ACM} Symp.\ on Principles of
  Programming Languages, {POPL}'90, San Francisco, {CA}, {USA}, 17--19 Jan
  1990}, pages 47--57. ACM Press, New York, 1990.

\bibitem{HofStr97}
Martin Hofmann and Thomas Streicher.
\newblock Continuation models are universal for lambda-mu-calculus.
\newblock In {\em LICS '97: Proceedings of the 12th Annual IEEE Symposium on
  Logic in Computer Science}, page 387, Washington, DC, USA, 1997. IEEE
  Computer Society.

\bibitem{Oga02}
Ichiro Ogata.
\newblock A proof theoretical account of continuation passing style.
\newblock In {\em CSL '02: Proceedings of the 16th International Workshop and
  11th Annual Conference of the EACSL on Computer Science Logic}, pages
  490--505, London, UK, 2002. Springer-Verlag.

\bibitem{ong97curryhoward}
C.-H.~Luke Ong and Charles~A. Stewart.
\newblock A {C}urry-{H}oward foundation for functional computation with
  control.
\newblock In {\em Conf.\ Record 24th {ACM} {SIGPLAN}-{SIGACT} Symp.\ on
  Principles of Programming Languages, {POPL}'97, Paris, France, 15--17 Jan.\
  1997}, pages 215--227. ACM Press, New York, 1997.

\bibitem{Par92}
Michel Parigot.
\newblock {$\lambda\mu$}-calculus: an algorithmic interpretation of classical
  natural deduction.
\newblock In {\em Logic programming and automated reasoning (St.\ Petersburg,
  1992)}, volume 624 of {\em Lecture Notes in Comput. Sci.}, pages 190--201.
  Springer, Berlin, 1992.

\bibitem{Par97}
Michel Parigot.
\newblock Proofs of strong normalisation for second order classical natural
  deduction.
\newblock {\em J. Symbolic Logic}, 62(4):1461--1479, 1997.

\bibitem{Plotk75}
Gordon~D. Plotkin.
\newblock Call-by-name, call-by-value and the lambda-calculus.
\newblock {\em Theor. Comput. Sci.}, 1(2):125--159, 1975.

\bibitem{SabFell93}
A.~Sabry and M.~Felleisen.
\newblock Reasoning about programs in continuation-passing style.
\newblock {\em Lisp and Symbolic Computation}, 6:289--360, 1993.

\bibitem{Sel01}
Peter Selinger.
\newblock Control categories and duality: on the categorical semantics of the
  lambda-mu calculus.
\newblock {\em Math. Structures Comput. Sci.}, 11(2):207--260, 2001.

\bibitem{StrReus98}
Th. Streicher and B.~Reus.
\newblock Classical logic, continuation semantics and abstract machines.
\newblock {\em J. Funct. Programming}, 8(6):543--572, 1998.

\end{thebibliography}
\bibliographystyle{plain}   

\end{document}